\documentclass[preprint]{aastex}

\def\stacksymbols #1#2#3#4{\def\theguybelow{#2}
    \def\vp{\lower#3pt}
    \def\sp{\baselineskip0pt\lineskip#4pt}
    \mathrel{\mathpalette\intermediary#1}}

\def\intermediary#1#2{\vp\vbox{\sp
     \everycr={}\tabskip0pt
     \halign{$\mathsurround0pt#1\hfil##\hfil$\crcr#2\crcr
              \theguybelow\crcr}}}

\def\gapproxeq{\stacksymbols{>}{\sim}{2.5}{.2}}

\shorttitle{Visual companions to close binary stars.}

\shortauthors{Rucinski \& et al.}

\begin{document}

\title{Contact binaries with additional components.~III.\\
A search using adaptive optics.\footnote{Based on 
observations obtained at the Canada-France-Hawaii 
Telescope (CFHT) which is operated by the National Research 
Council of Canada, the Institut National des Sciences de 
l'Univers of the Centre National de la Recherche Scientifique 
of France, and the University of Hawaii. 
}}

\author{Slavek M. Rucinski\altaffilmark{2},
Theodor Pribulla\altaffilmark{3},
Marten H. van Kerkwijk\altaffilmark{2}}

\altaffiltext{2}
{Department of Astronomy, University of Toronto,
50 St.~George St., Toronto, Ontario, Canada M5S~3H4;
(rucinski,mhvk)@astro.utoronto.ca}

\altaffiltext{3}
{Astronomical Institute, Slovak Academy of Sciences,
059~60 Tatransk\'a Lomnica, Slovakia;
pribulla@ta3.sk}

\begin{abstract}
We present results of the CFHT adaptive optics search for
companions of a homogeneous group of contact binary stars, as a
contribution to our attempts to prove a hypothesis that
these binaries require a third star to become so
close as observed. In addition
to companions directly discovered at separations of
$\ge 1''$, we introduced a new method
of AO image analysis utilizing distortions
of the AO diffraction ring pattern at separations
of $0.07'' - 1''$.  Very close companions,
with separations in the latter range
were discovered in systems HV~Aqr,
OO~Aql, CK~Boo, XY~Leo, BE~Scl, and RZ~Tau. More distant
companions were detected in V402~Aur, AO~Cam, V2082~Cyg.
Our results provide a contribution to the mounting evidence
that the presence of close companions is a very common phenomenon
for very close binaries with orbital periods $<1$ day.
\end{abstract}

\keywords{ stars: close binaries - stars: eclipsing binaries --
stars: variable stars}

\section{INTRODUCTION}
\label{intro}

The formation of close binaries is still a puzzle: How can they be formed
if the components were larger than the instantaneous orbit during the
pre-main-sequence phase? One possibility is that angular
momentum transfer to a third companion caused the originally
wider orbit to shrink during the main-sequence (MS) phase. 
If this hypothesis is true, all close binaries should 
be accompanied by a third body (unless it was removed by an encounter, 
but this is unlikely by the time the stars have reached the MS).  Thus,
one would expect the incidence of the triples
to be higher than in a sample of relatively wide (P$\gapproxeq$10
days) binaries. 
\citet{tokov2002} and \citet{tokov2006} studied the
frequency of occurrence of tertiary companions for the binary
period range of 1 to 30 days and collected an impressive amount of
evidence that indeed the frequency of tertiaries increases as the binary period
decreases, from a level of about 34\% for orbital periods
of 12 -- 30 days to 50\% at 9 days, and further to perhaps even 100\%
at 1 day.

This paper is the third in a series assessing the presence and
properties of tertiary companions to very close binaries
with periods shorter than 1 day. 
In the first paper of the series, \citet[ Paper~I]{priruc06},
we summarized the present state, collecting
all detections of third and/or multiple
components to contact binaries.  This led to a heterogeneous sample,
with observational biases which were difficult to quantify.
Nevertheless, for our sample of contact binaries brighter
than $V_{\rm max} = 10$, it allowed us to set a firm lower limit of
$59 \pm 8$\% to the incidence of the tertiaries (for the
better-observed sub-sample of the Northern hemisphere
systems). The preliminary results of adaptive optics program
discussed in the current paper were reported in Paper~I.

In \citet[ Paper~II]{dang06}, we presented 
a spectroscopic search for faint third components.
Several thousand medium-resolution spectra obtained during the David
Dunlap Observatory (DDO) radial velocity program for 80 binaries were
re-analyzed, and weak -- but stable -- spectral signatures 
of tertiaries were searched for in the averaged spectra. This resulted in a
detection of 15 tertiaries -- of which 11 were previously unknown --
in a homogeneous sample of 59 contact binaries. The continuation of
the radial velocity observations -- 110 good accuracy orbits
are now published \citep{ddo12} -- will provide material for a
future similar study. The spectroscopic observations 
permit detection of very close companions 
hidden in the seeing disk of the close binary
(at DDO typically 1.8 -- 2 arcsec) but -- for faint companions --
it is limited to a magnitude difference of 
about 5 magnitudes (e.g., in CK~Boo, $L_3/(L_1+L_2) = 0.009$, Paper~II).
The AO observations presented in the current paper   
permit detection of much fainter and redder companions, 
particularly in the infrared, but only at relatively large separations,
and with long-period orbits, compared to those studied spectroscopically.
Unlike for spectroscopic observations, where a physical
bond can often be proven by similar systemic velocities
or mutual revolution, visual detections for nearby ($<300$ pc)
and relatively bright ($V <10$) binaries may suffer
some contamination from projections; often, however, the density of
background stars is sufficiently low that the mere presence
of an object at small separation from a bright star is a strong
indication of a physical bond.

The majority of the systems covered in this series of papers are genuine contact
binaries. At orbital periods shorter that one day they are the most
frequently detected and studied. However, our program includes also
binaries with unequally deep eclipses which may be contact binaries
with poor thermal contact or short-period semi-detached systems. 
In some cases (V1464~Aql, TT~Cet), low photometric amplitudes 
and/or lack of spectroscopic observations preclude any meaningful
classification. We note, that -- evolutionary -- 
all these types must be related to contact binaries as binaries of the
least orbital angular momentum. Hence, we will interchangeably use
the terms ``close'' and ``contact'' binary stars throughout 
this paper to mean all target binaries with periods shorter than 1 day.

This paper is structured in the following way: In Section~\ref{obs}, we
describe our observations and their reduction, and in 
Section~\ref{close}, we introduce a new search technique for
multiples, which uses template fitting, and present our new detections.
In Section~\ref{limits}, we use a Monte-Carlo technique to verify
our estimate of the detection limits for a given magnitude
difference.  In Section~\ref{bond}, we present tests of the 
physical bond between the visual companions, and in
Section~\ref{indiv}, we discuss the nature of the detected
companions. Finally, in Section~\ref{summary}, we summarize and
interpret our results.

\section{OBSERVATIONS}
\label{obs}

We obtained observations aimed at the direct detection of companions of
contact binary stars on the nights of 1998 January 10 and 1998 July 23, 
2005 October 17 and 18,
using the adaptive optics (AO) system at the Canada France Hawaii
Telescope (CFHT). During the 2005 run, the specific goal was to survey the 
many bright contact binaries originally detected by the Hipparcos 
satellite and later confirmed spectroscopically during radial velocity 
programs (mostly at the David Dunlap Observatory; for reference, see
\citealt{ddo11,ddo12}). 
Altogether, 80 known contact binaries stars -- accessible from CFHT 
to an approximate limit of $V <10.5$ -- were observed with 15 objects 
observed on two occasions and with 2 objects observed three times. The 
journal of all observations can be found in Table~\ref{tab1} and overview
of the detections is given in Table~\ref{tab2}.

We used the PUEO AO system with the KIR camera combination
\citep{rig1998}, with a $1024\times1024$ pixel detector that covers a
field of $36 \times 36$ arcsec at a scale of
0.0348${\rm\,arcsec\,pixel^{-1}}$.  In 1998, most observations were 
done in the infrared $K$ band or, to avoid overexposure of the contact
binary, a narrow-band filter centered on the CO bands ($K_{CO}$; for
details on the filters, see notes to Table~\ref{tab1}); a few binaries  
were also observed in the $H$ band (1.65 $\mu$m).
In 2005,  we used the $H_2$ filter, which for our purposes is
effectively another narrow-band $K$ filter.

In 1998, each source was observed by positioning it in each of the
four quadrants of the detector, while in 2005 a fifth,
central position was added. 
At each position, typically three to five exposures
with integration times from 5 to 40~s were obtained; typically, we
thus collected 12 to 25 individual
images.  With the image shifts and multiple exposures, we could obtain
good definitions of the background and -- for detected 
visual companions -- reliable uncertainty estimates of magnitude
differences ($\Delta m$), separations ($\rho$) and position angles
(PA, $\theta$).

In all cases, the contact binaries themselves were sufficiently bright
to be used for sensing the distortion of the wavefront and deriving
the AO corrections.  In the $K$ passband, the isoplanatic
patch is relatively large, about $40''$, and thus encompasses the
whole field of view of the KIR camera. The Strehl ratio of the
AO-corrected images was found to be almost constant (around 0.30) for
all images taken in natural seeing better than 0.9 arcsec. On a few 
occasions of a worse seeing, the Strehl ratio fell below~0.2. 
Although the amount of light scattered from the Airy disk into the 
rings increased on such occasions, the diffraction pattern was still 
very well defined. The measured FWHM of the AO-corrected images was 
0.143 arcsec in the $K$ band, close to the expected diffraction-limited 
performance of a 3.6 m telescope.

The resulting point spread function (PSF) of the corrected images
is a combination of a diffraction pattern defined by the aperture
of the telescope with a residual Gaussian and/or Lorenzian profile 
of much less understood character. Its shape was found to vary 
depending on the declination and the hour angle of the telescope 
(see Section~\ref{close}).

For the initial reductions, we subtracted bias and dark current from the raw
images, divided by flat fields, and removed hot pixels.  
Next, we co-added all images for different sky displacements.
For astrometry, we measured the rotation and scaling of the
AO system and its detector using known wide visual pairs with very slow
orbital motion. The position angles for the 1998 and 2005 observing runs
had to be corrected by $-1\degr$ and $+4\degr$, respectively, but no 
corrections to the pixel scale were needed. For both 1998 runs the 
correction was found to be the same.

\section{DETECTION OF VISUAL COMPANIONS}
\label{close}

In order to obtain maximum sensitivity to faint, close companions, we
attempted to model the observed PSF of single targets.  We first 
used an analytical combination of the Bessell function
of the first order for the diffraction-limited part and 
a Gaussian function for the residual
uncorrected light. The resulting representations turned out
to be poor due to additional asymmetries in the PSF,
visible as ``gaps'' at various position angles in the diffraction rings.
Clearly, a heuristic approach was needed which
would utilize the observed PSF shape as given by this particular
combination of telescope, AO system, and camera.

\begin{figure} 
\plotone{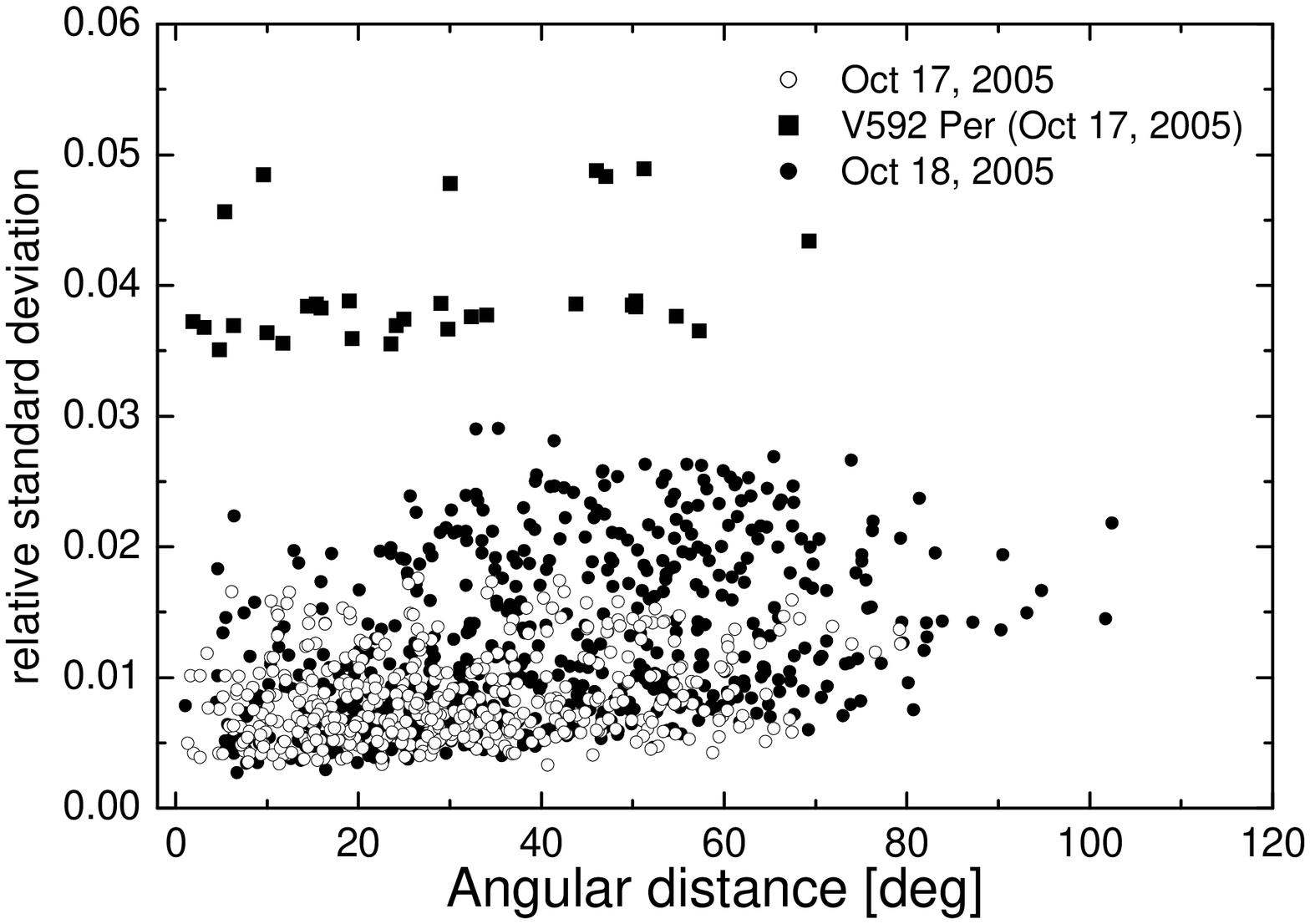}
\caption{Similarity of the point-spread functions as a function of the
angular distance on the sky.  The relative standard deviation of mutual 
fits for all possible pairs of stellar
images.  The images were matched within each night separately, as
marked by different symbols.  The system V592~Per stands out, as
expected given that it has a very close, relatively bright
companion. 
\label{fig1}}
\end{figure}

\begin{figure} 
\plotone{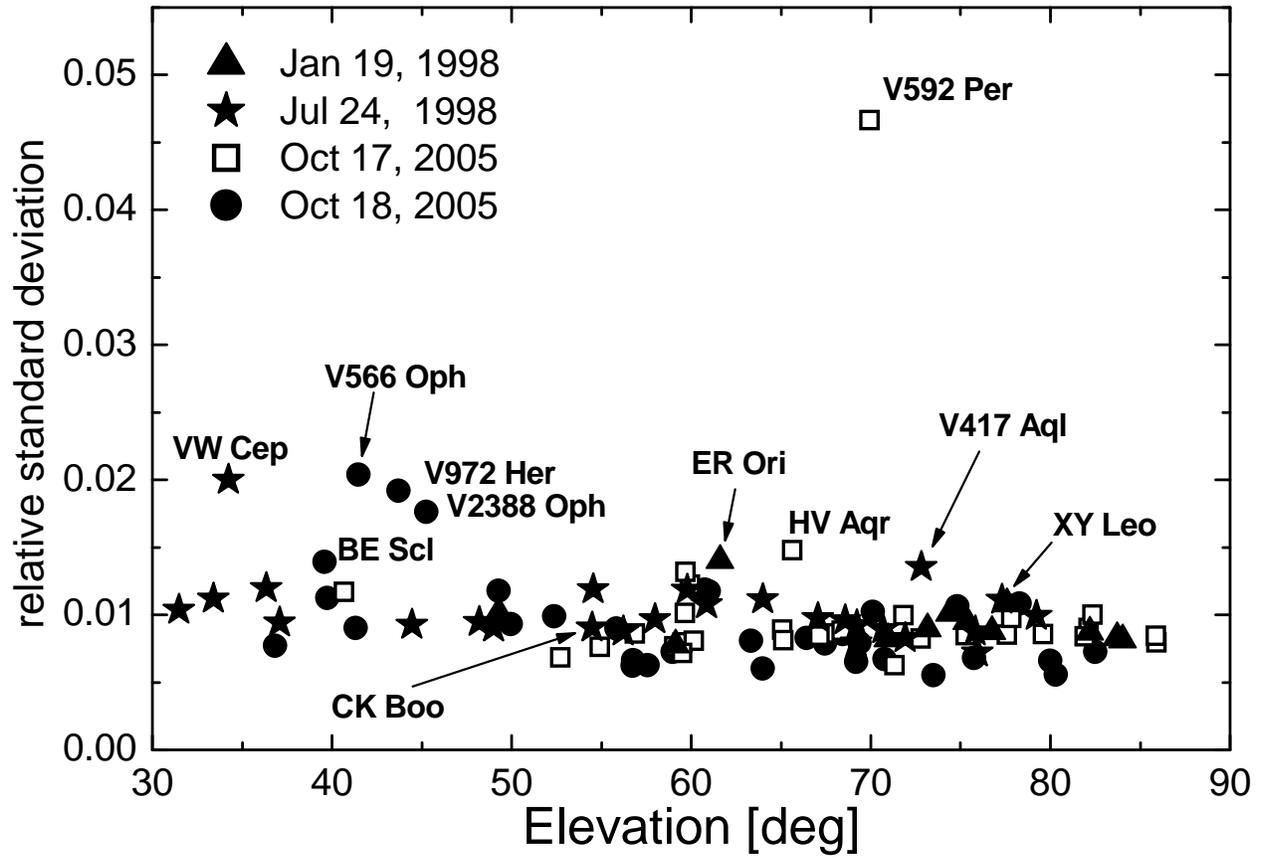}
\caption{Similarity of the point-spread functions as a function of
  elevation.  The  relative standard deviation is computed for
  all templates used to fit PSF of a given object. Objects showing large
  mismatches are suspected to have companions. 
\label{fig2}}
\end{figure}

Careful inspection of our images revealed that
the PSF shapes were similar for most of the stars,
but contained small residual deformations which depended
primarily on the position of the telescope 
(Figures~\ref{fig1} and \ref{fig2}) and -- to a lesser
degree -- on the instantaneous seeing and the effective Strehl
ratio of the corrected image. To investigate
the variation in the PSF shape, we fitted each stellar image to all
images of other stars. 
While most systems showed just mild dependencies of the similarity
in the image shape on various positional factors,
some had systematically different shape.
We quantified the differences by a relative (to the
maximum within the image) root mean square standard deviation
of all fits to a given object using other images as templates.
Some of the systems showing large mismatches are well known
close visual binaries where the secondary is located at very small
separation (VW~Cep, V2388~Oph, ER~Ori, V592~Per), while some others
were previously unrecognized as visual binaries (XY~Leo, BE~Scl). 
These results thus led to a cursory, but rapid  
identification of close pairs.

\begin{figure} 
\plotone{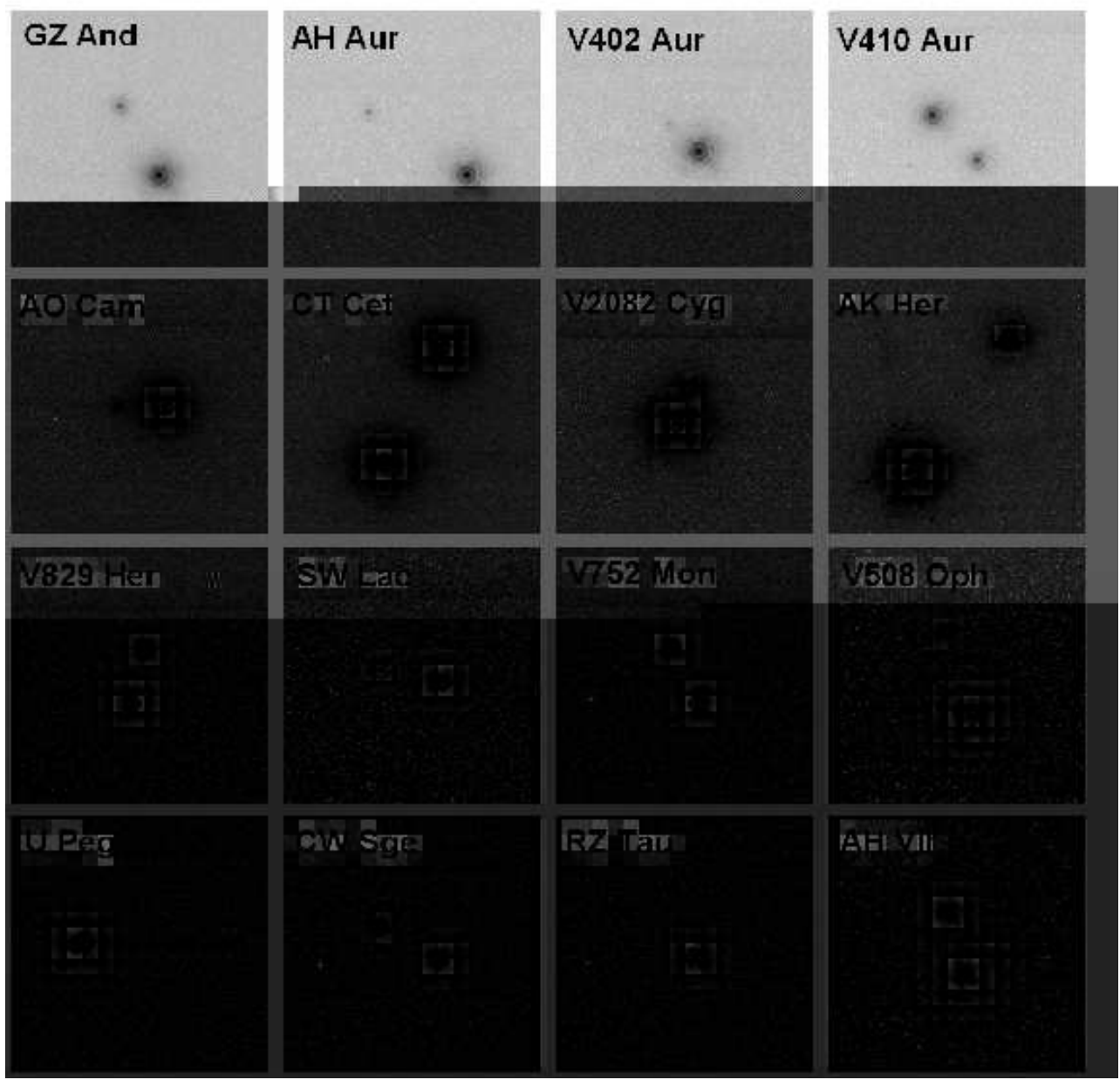}
\caption{A mosaic showing systems for which the presence of a
  companion was obvious. The width of each panel corresponds to 7
  arcsec. \label{fig3}}
\end{figure}

\begin{figure} 
\plotone{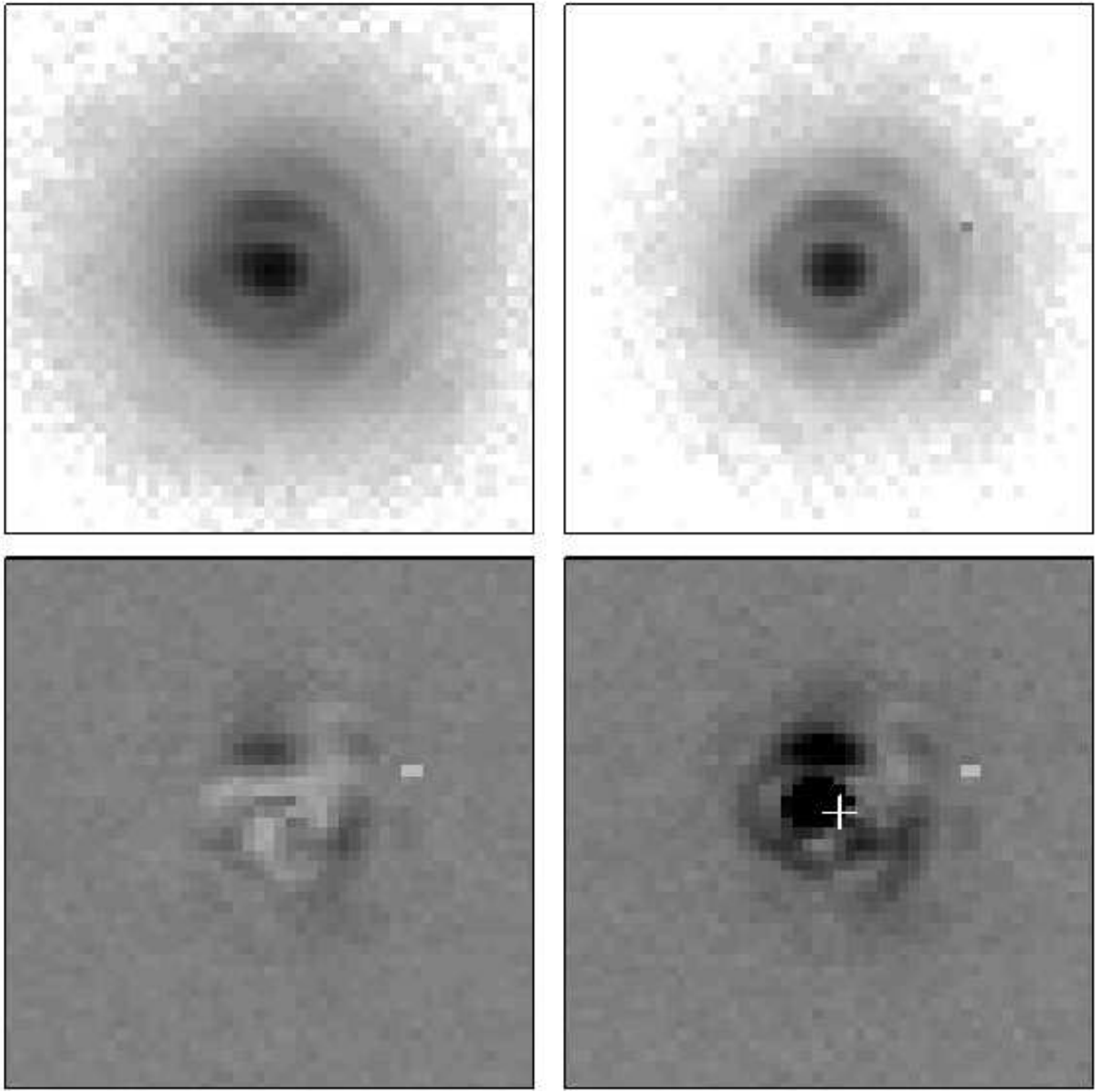}
\caption{The process of fitting a model binary using a template to the
very close visual binary BE~Scl. The four panels show, from the top
left, clockwise: (1)~the object image, (2)~the template, (3)~the
secondary component after subtraction of the primary 
(its location marked by a white cross) and (4)~the
residuals from the best fit. We used the five individual images
(obtained by stacking several exposures at a given quadrant
displacement) and fitted them by the best single exposures from
46 templates.  Of the resulting 230 fits, 211 succeeded, while 19 did
not; the figure shows one of the successful ones.  Similar figures
for the eight remaining very close systems are available in the electronic
version of the paper only.  In
this and the following figures, the images are oriented with North up
and East left. \label{fig4}}
\end{figure}

Following this preparatory stage, we analysed the images in more
detail in two steps (i)~we determined positions and magnitude
differences of obvious and distant ($\rho > 1.0$ arcsec) companions from 
fluxes and centroid positions of their Airy disks (the mosaic of 
direct detections is presented in Fig.~\ref{fig3}); and (ii)~we
performed an automated search for faint companions which were missed
in visual inspections or were too close to measure their positions 
and brightness.
For the latter search, we made direct fits to a binary model assuming
that -- within the isoplanatic patch --
a binary can be represented by two properly
shifted and scaled images of a well-chosen template. 
Thus, we express the observed image intensity for a visual double,
$I(x,y)$ as,
\begin{equation}
I(x,y) = B + A_1 * T(x-x_1, y-y_1) + A_2 * T(x-x_2, y-y_2),
\end{equation}
where $B$ is the background level, $A_1$ and $A_2$ are
the normalization factors of the two stellar images (central
star and companion, if present), $T(x,y)$ is the  template image, and 
$x_1,y_1$ and $x_2,y_2$ are the spatial shifts of the template required to
match the fitted images (for non-integer pixel shifts, we use
bi-linear interpolation in the template intensity). Thus, the fits
involve three parameters entering linearly and four parameters
entering non-linearly. The performance of our technique is illustrated
in Fig.~\ref{fig4} for a very close companion to 
BE~Scl\footnote{Similar figures for eight additional systems, 
OO~Aql, HV~Aqr, CK~Boo, VW~Cep, XY~Leo, V2388~Oph, ER~Ori, V592~Per, 
are available in the electronic version only, see the
caption to Figure~\ref{fig4}.}.

We found that the largest difficulty in identifying
faint companions which might create only slight deformations of the
PSF, was to find a suitable template.  To be as objective as possible,
we performed an automated, brute-force search in which we fitted
each object by a visual binary model which uses, in succession, all
other objects observed in a given run as templates (but omitting
V592~Per and other objects already recognized as close pairs).  To
ensure the fitting process did not converge only to local minima,
and to increase chance of finding real faint/close visual companion(s),
we chose 24 initial positions of potential companions: eight different
position angles $\theta$, in steps of 45$\degr$, for three different
initial separations $\rho$, of 0.1, 0.2, and 0.3 arcsec,
(i.e., sampling the region from the core to just outside the
diffraction ring ). As a starting magnitude difference, we used 
$\Delta M = 3$~mag.  For a given object and a template combination, the 
fit with the best statistical quality, $\chi^2$, from all possible 
starting sets of parameters was recorded. We rejected any fits that 
converged to separations closer than 0.02 arcsec.

Using as many templates as possible (e.g., 46 different stars for the
2005 observing run), we were able to assess the reliability of the
detections and their significance level from the distribution of the
recovered parameters.  Also, with as many as 200 resulting parameter
sets for a given object, we could reliably determine final (mean)
values and uncertainties for the various parameters for positive
detections. For the 1998 runs, we omitted a small number of
$H$ and $K$ band observations and utilized only frames taken in the most
frequently used filter, $K_{\rm CO}$.

\begin{figure} 
\epsscale{0.60}
\plotone{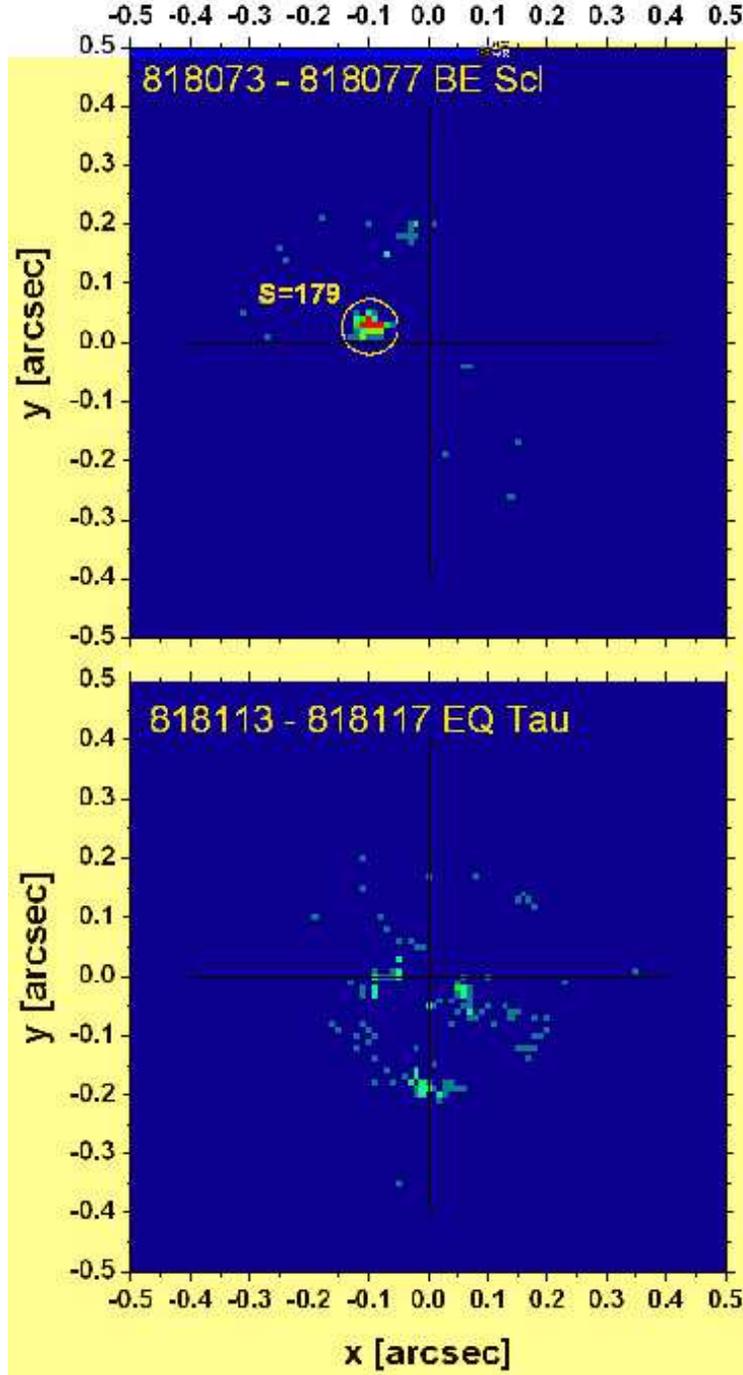}
\caption{Convergence count maps from the automatic search routine (as
described in the text) for BE~Scl and EQ~Tau, the cases 
where images did not show any obvious 
evidence of a companion, but the automatic search showed a 
clear detection for BE~Scl, and no companion for EQ~Tau.  
The total number per pixel is indicated by color. The circle for
BE~Scl encompasses a tight cluster of counts, in which 179 out of 211
of the fits indicated a companion could be present, indicating a real
detection. No similar clustering is seen for EQ~Tau. The 
maximum number of counts per pixel anywhere in the image is 
4 for EQ~Tau, but 52 for BE~Scl. Similar figures
for all systems analyzed with the automatic search routines
are available in the electronic version of the paper.  
\label{fig5}}
\end{figure}

One of the main difficulties in analyses like ours is to determine the
significance of detections.  We assessed it using the following scheme:
For each successful convergence of the secondary position to a pixel
inside an area of $100 \times 100$ pixels around the primary, we added a 
convergence unit ``count'' to that pixel, and we then looked in these
``convergence maps'' for count clustering. In other words, 
we took high numbers of
counts for {\it all templates\/} as an indication of a 
distortion in the image which could be due to the presence of a
companion.  With this method, one immediately recovers close, but
obvious pairs, for which almost all object-template fits led to
consistent sets of parameters.  But one can also recover less obvious
cases, as is illustrated in Fig.~\ref{fig5}\footnote{The
convergence count maps for all objects are available at
http://www.astro.utoronto.ca/~rucinski/Triples3/.}
where we show convergence maps for two systems, EQ~Tau and
BE~Scl.  Here, EQ~Tau is a typical case of no detection with a few
spurious instances of the algorithm locking on small fluctuations within 
the first diffraction ring. However, for BE~Scl, one sees a clear 
clustering of successful convergences within a small circular area 
indicating a genuine detection.

\begin{figure} 
\epsscale{1.00}
\plotone{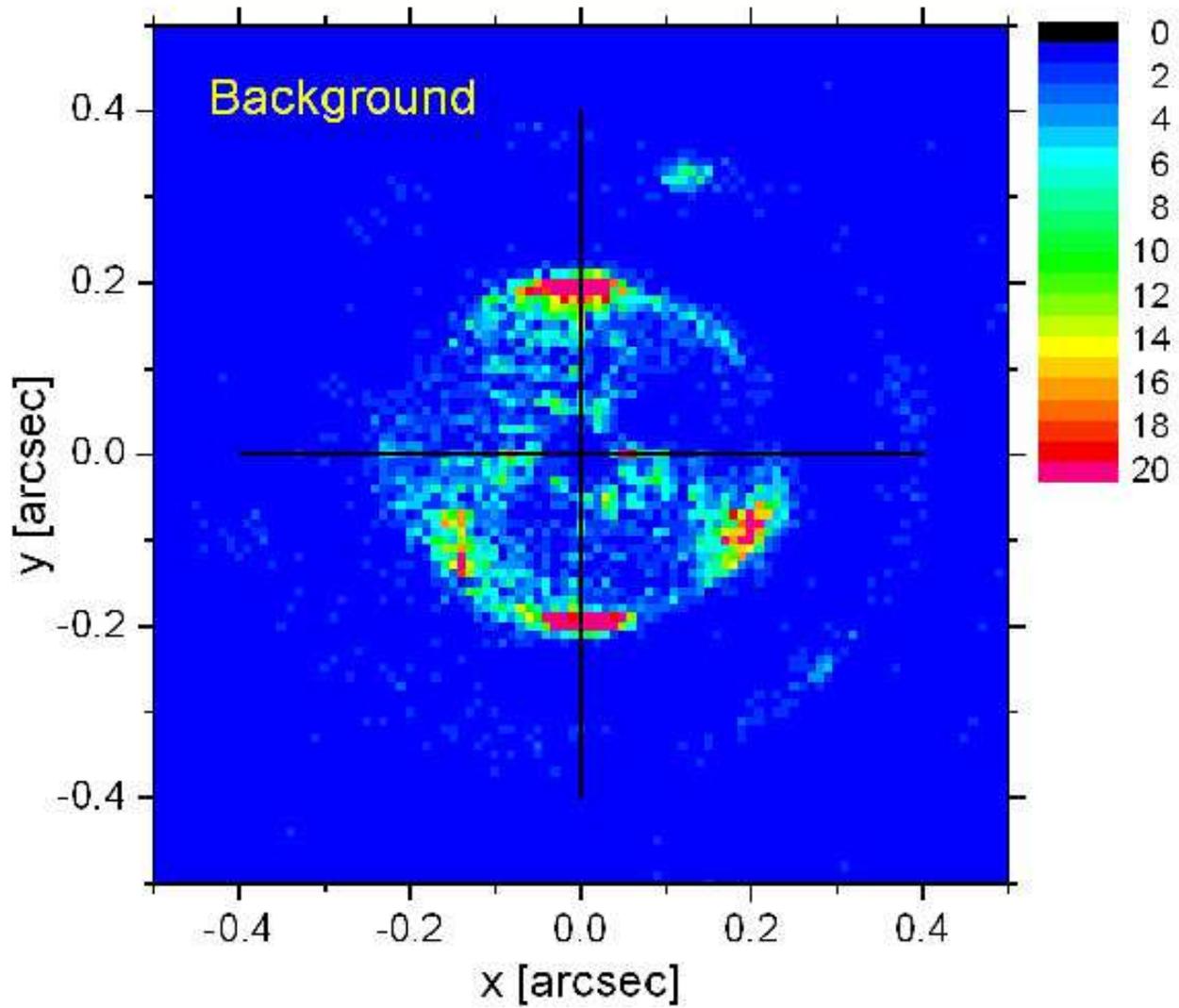}
\caption{The background of false detections for all stars which do not
show any visual companions (as observed in 2005; the 1998 images are
very similar). The first diffraction ring and traces of the second
ring are clearly visible. \label{fig6}}
\end{figure}

The approach described above has a limitation related to the
increased random fluctuation noise in brighter portions of the image, 
particularly within the first ring, where the likelihood of the
algorithm locking on random positive deviations is much larger. 
To assess the probability of false detections, 
we constructed a map similar to those we made for the individual stars, 
but obtained by adding the convergence maps for all stars which showed no
indication of a companion (as for EQ~Tau in Fig.~\ref{fig5}).
In practice, we added the low numbers of false detections in the
diffraction rings for individual stars and obtained a 
``background map'' of false detections against which our detections
can be compared (Fig.~\ref{fig6}).  Note that we constructed these
background images separately for 1998 and 2005 runs; encouragingly,
though, the images are very similar, indicating the characteristics of
the CFHT AO system did not change much between these two epochs. 

With the convergence and background maps in hand, we assigned
significance of the detection using two approaches, (i)~by simply taking 
the count maximum within the map, or (ii)~by integrating (summing)
the counts within a small (e.g., $r < 6$ pixel) aperture and comparing 
this with the number of counts in the same region in the background map.  
For both approaches, the relevant quantity required normalization by the 
number of used object and template frames.  For each object and observing
run, Table~\ref{tab3} gives number of the object frames, $N_f$, and
the number of templates used, $N_t$ (28 in 1998 and 46 in 2005), and 
the number of frames used to construct the background map, $N_b$.
For the first approach, the maximum number of counts within the count 
map ($C$) was then normalized to a typical number of object frames
and templates of the 2005 run ($N_f$ = 5, $N_t$ = 46) as:
\begin{equation}
C_n = C \frac{230}{N_f N_t}.
\end{equation}
In the second approach, the sum of counts $S$ in a cluster within a
given aperture of radius $r$ is divided
by the corresponding summed count number (within the same area)
in the background image $B$. The quality of the detection is
then expressed as the normalized ratio of counts in the object and the
background count maps:
\begin{equation}
R_n = \frac{S}{B} \frac{N_b}{N_f}
\end{equation}

In Table~\ref{tab3}, we list the systems that show clustered counts
within small circular apertures (with $r < 6$ pixels), and are thus
suspected higher-order systems, while in Table~\ref{tab4}, we give
maximum normalized counts $C_n$ for systems for which no companions
are suspected.  Selection of a threshold for actual detection based on
the ``quantities of merit'', $C_n$ or $R_n$, was rather difficult and
somewhat arbitrary; after some experiments, we felt a best distinction
was achieved by taking as detections those objects that had $C_n > 20$
or $R_n > 50$, as listed in the last column of Table~\ref{tab3}.
With these thresholds, the automated search confirmed all suspected
cases seen as deformation of the diffraction structure. 
These thresholds lead to the confirmation of the four known tight
pairs -- V592~Per, VW~Cep, ER~Ori (1998 run) and V2388~Oph -- and four
new detections -- HV~Aqr (this one is also directly visible in the
image), BE~Scl, XY~Leo and CK~Boo.  Among these, XY~Leo is the closest
resolved system, with a separation of only 0.061 arcsec. The suspected
cases -- V376~And, TZ~Boo, DN~Cam, YY~Eri, BV~Eri, V829~Her, V508~Oph,
V351~Peg, and TY~Pup -- clearly require new observations. The results
for the automated search (presented in Tables~\ref{tab3} - \ref{tab4})
apply only for the detection of very close companions within 0.3-0.4 
arcsec. Hence, e.g., V829~Her having distant companion clearly visible
at 1.46 arcsec is given as detection ``D'' in Table~\ref{tab2} but
there is only suspicion for very close companion (Tables~\ref{tab3} 
- \ref{tab4}).

Measured properties for the detected companions are given in
Table~\ref{tab5}.  The table gives designations of the components, the
separation ($\rho$), the position angle ($\theta$) in degrees, the
magnitude differences in $K$ and $H$ bandpasses, the heliocentric Julian
date of observation and, from the date, the orbital phase of the
binary and a correction of its brightness (see Section \ref{companions}).
The table lists only systems with detected companions within 5 arcsec.
Although this limit seems to be set arbitrarily, even faint companions
separated by more than 5 arcsec can be easily detected with small telescopes
without adaptive optics. A good example is GZ~And, where we give only our
new detection at 2.13 arcsec; the components B and C,
separated by 8.6 and 13.3 arcsec, respectively, are clearly
visible even in small telescopes \citep{walk73}.

\section{EVALUATION OF DETECTION LIMITS}
\label{limits}

The limit to which we were able to detect tertiaries for a given
separation depended mainly on (i)~the magnitude difference of the
components in the $K$-band, $\Delta K$, and (ii)~the similarity of
the PSF for the object and the template.
We assessed these limits by adding artificial stars to our real
frames and applying the whole detection procedures. 

We selected images of two single stars without 
any close ($\rho < 10$ arcsec) companions: AQ~Psc serving as an object
and FP~Eri as a template. Artificial images were produced by shifting and
co-adding the same object image for a wide range of separations
($0.04 < \rho < 1.20$ arcsec) and magnitude differences ($0 < \Delta K
< 6.5$ mag).  
For given $\Delta K$, the detection limit was determined as a minimum
separation for which the algorithm would converge
to the original position to within the FWHM of the PSF (we found our
result was insensitive to the exact choice of convergence criterion). 
The resulting detection threshold curve, together with all actual detections, 
is displayed in Fig.~\ref{fig7}; its ruggedness reflects the complex 
diffraction pattern in the PSF.  This curve is very similar to
curves published before for the same system \citep{duch1999,bouv2001}.  
As one can see in Fig.~\ref{fig7}, the above criterion for the
detection limit derived from artificial stars is consistent with what we
found from our automated search: all detected systems are above or at
the detection limits.

\section{NATURE OF THE COMPANIONS}
\label{companions}

We attempted to estimate nature of the companions
-- assuming they are physically bound -- from the measured brightness
differences and estimates of the absolute magnitude of the close
binaries.  The results are listed in Tables~\ref{tab5} and~\ref{tab6},
with the former giving results for individual observations and the
latter overall properties that we used or determined. 

To obtain the best constraints on the companions, we first needed to
ensure our magnitude differences were evaluated relative to the
maximum brightness of the contact binary.  This is important for
some contact binaries showing large photometric amplitudes (e.g.,
GZ~And, OO~Aql, SW~Lac, ER~Ori, and U~Peg with $\Delta V > 0.70$ mag).
Thus, knowledge of the orbital phase is necessary for a proper
estimate of the instantaneous brightness of the contact binary.  The
phases at the time of the AO observations were calculated using the
Cracow on-line, up-to-date ephemeride
database\footnote{http://www.as.ap.krakow.pl/ephem/}
\citep{kreiner01,kreiner04}.  The visual magnitude of the contact
binary was then estimated by using published light curves (sometimes
by graphical tracing in figures) to determine the difference with the
maximum brightness $V_{\rm max}$ (the latter usually taken from the
General Catalogue of Variable stars\footnote{We used the most recent
electronic version (4.2), available at
http://www.sai.msu.su/groups/cluster/gcvs/}).

To convert these $V$ magnitude at the time of the observation to
absolute magnitudes $M_V$, we used Hipparcos parallaxes with relative
precision better than 15\% if available (an error of $<\!0.30$ mag
in the absolute magnitude), and the period-color-luminosity
relation of \citet{rd97} (an error of $\sim\!0.30\,$mag) for
systems with poorer parallaxes (GZ~And, V829~Her, OO~Aql,
AO~Cam, and ER~Ori).  Because all objects are nearby, we neglected
interstellar absorption.

Next, to calculate expected absolute magnitudes $M_K$ at the time
of the observation, we used main-sequence colors appropriate for the
spectral types of the binaries (as determined from DDO observations;
see \citet{ddo11,ddo12} for references to previous publications). 
Here, we note that in comparing contact binaries with single stars,
one must take into account that while they are MS objects, the energy
transfer from the primary to the secondary component makes the primary
component always cooler than a MS star of the same mass \citep{mochna81}.
Furthermore, because of the increased radiating area, the absolute
visual magnitude $M_V$ of a contact binary obtained directly from its
trigonometric parallax and $V_{max}$ is always brighter than the $M_V$
corresponding to a single MS star of the same spectral type.  However,
for given spectral type, contact binaries appear to have the same
colors (S.M.R., unpublished comparison of $B-V$ and $V-K$ indices),
so that our simple corrections are adequate to evaluate $M_K$.  

Finally, with $M_K$ for the contact binary at the time of the
observation, the observed magnitude difference then yields the
absolute magnitude of the companion, and, with the appropriate
main-sequence relations, the estimated spectral type.  One
sees that most of the companions appear to be late, M0--M6 dwarfs,
with some, such as those of U~Peg and V402~Aur, of even later spectral
types. 

To estimate orbital periods of wide visual orbits of triple systems, we 
also calculated the total masses of the contact binaries, by combining 
radial-velocity orbits such as those obtained at DDO \citep{ddo11,ddo12}, 
which give $(M_1+M_2) \sin^3 i$, with photometric observations, which 
provide the orbital inclination $i$.  In the selection of such combined 
solutions, we preferred the total masses determined by simultaneous light and
radial-velocity curve fits \citep{gaze06}. The total mass of the systems 
(using estimated mass of third companion) was then used to estimate an 
approximate orbital period of the visual pair from the angular separation 
and the distance.

\section{THE PHYSICAL BOND}
\label{bond}

While the presence of a star close to several
contact binaries is unquestionable, the physical link is
often difficult to prove. There are three possible ways of
assessing the physical bond:

\paragraph{Observed color difference.}  
With observations in multiple photometric bands, 
one can verify that the spectral type of the
companion inferred from its magnitude -- assuming the same distance as
the contact binary -- is consistent with its colors.  Unfortunately, we
can do this for few systems only, since most of the data were obtained
in one passband ($K$ and its narrow-band substitute).  Moreover, even
for systems observed in $H$ and $K$, the constraint is weak, since the
$(H-K)$ color index is insensitive to the spectral type \citep{cox00}:
For most MS stars, it is in the range $-0.04 < H-K <0.11$ and starts
increasing to $\simeq 0.33$ only at spectral types of about M5V.
We find that in our current sample,
the available color indices usually agree moderately well with the
absolute $K$-band magnitudes of the companions (see
Section~\ref{indiv}).  For future confirmations, a more suitable index
would be the $(J-K)$ color index, which increases monotonously from
$-0.22$ at O5 to 0.86 at M2, then becomes nearly constant in M dwarfs,
again to increase strongly for L dwarfs.  A disadvantage of $(J-K)$,
though, is that the performance of AO systems is substantially worse
in the $J$ passband.

For some binaries, we can compare the infrared magnitude differences
with the visual flux ratio $L_3/(L_1+L_2)$ obtained from the
averaged spectra.  This gives a much larger wavelength base, but works only
in cases when the third component was located well within the
spectrograph slit.  For consistency, the visual flux ratio should be
measured relative to maximum light of the contact binary.

\paragraph{Similarity of the proper motion.}
For systems where the contact binary has a relatively high proper
motion, say 
$\mu=\sqrt{\mu_\alpha^2 \cos^2 \delta + \mu_\delta^2}\,>50$ 
mas~yr$^{-1}$, an {\it optical\/} companion would show an
optical projection motion between our 1998 and 2005 observing runs,
while a physically bound one would not (provided the orbital period is
sufficiently long that no orbital motion is expected). 
To test this, we collected proper motions (Table~\ref{tab6}) from the
TYCHO~2 Catalogue \citep{tycho2} for all systems except GZ~And (which
was not observed, and for which we adapted the proper motion from the
UCAC2 Catalog; \citealt{ucac2}).  We found that the companions to the
large proper motion contact binaries AH~Aur, SW~Lac, U~Peg, and RZ~Tau
had the same relative position in both the 1998 and the 2005 runs, and
therefore for all four cases we thus are virtually certain that the
companions are physical ones.  

Among the systems observed only on one occasion, there are a few
further ones with large proper motion, for which the test should be
feasible in a few years.  These are HV~Aqr, V508~Oph, CK~Boo, and
V2082~Cyg (see Table~\ref{tab6}).

\paragraph{Similarity in radial velocities.}
Another kinematical indicator of physical association is a similarity of
the center-of-mass radial velocity of the contact binary and of the
radial velocity of the companion, a condition which should be valid
for companions on long-period orbits. Indeed, several companions
detected within the current program (GZ~And, HV~Aqr, CK~Boo, AO~Cam,
and V2082~Cyg), independently found from the analysis of the averaged
spectra from DDO in Paper~II, show radial velocity consistent with the
systemic velocity of the contact binary.

\section{RESULTS FOR INDIVIDUAL SYSTEMS}
\label{indiv}

We give summaries of the properties of the contact binaries with newly
detected companions in Table~\ref{tab6}. 
Here, we discuss some of the new detections, as well as systems 
with interesting non-detections. 
The WDS numbers below refer to entries in the Washington Double
Star Catalog \citep{wds}.

\begin{description}
\item[GZ~And.]  A new component found in the trapezoidal multiple
system GZ~And (WDS J02122+4440) is very probably a physical member:
the estimated $M_K$ indicates the spectral type M5V or later, and this
is supported by its color $(H-K) = 0.25$, which is appropriate for a
M3-4 dwarf.
\item[HV~Aqr.]  Indications of a faint tertiary to HV~Aqr were first
found in Paper II, with an estimated light contribution at 5184~\AA\
of $L_3/ (L_1 + L_2) \approx 0.022$ ($\Delta V = 4.14$ mag) and a
companion temperature of $T_3\simeq4000~$K.  This is consistent with
the infrared AO observations, which give $\Delta K = 2.00$ mag and
from which we inferred F0 and K5 spectral types.
\item[AH~Aur.]  Although the visual companion to AH~Aur is relatively
distant, at 3.16~arcsec, the proper motion indicates a physical
association: if the visual companion was a distant background star,
the relative position of components should change by
$\sim\!0.15$~arcsec between the two epochs of our observations, but it
has been found to be stable within 0.02~arcsec.
\item[CK~Boo.]  A late-type companion to CK~Boo was first found in
Paper~II, with $L_3/ (L_1 + L_2) = 0.009$ ($\Delta V = 5.11$) and
$T_3\simeq3900~$K.  This is consistent with what is expected for the
M0V dwarf inferred from our AO observations.  The detection in our
automated search seems to be particularly reliable because the
component appears in a practically empty part of the background count
map.
\item[VW~Cep.]  The companion to VW~Cep (WDS J20374+7536) was discovered
by \citet{hein74}.  The last published astrometric observation in the
WDS is from 1999 with $\rho=0.4$~arcsec and $\theta = 187\degr$.
Our position, $\rho$ = 0.254 arcsec and $\theta = 167\degr$, is closer
to the periastron passage which occurred in January 1997 and is
consistent with the position at $\rho = 0.249$~arcsec and $\theta =
165.2\degr$ expected from the elements in Sixth Catalog of Orbits of
Visual Binary Stars (see \citet{hart2001}). Since there appear to be
no other observations close to periastron, our new position has a potential
to markedly improve the companion orbit.
\item[CV~Cyg.]  From Hipparcos measurements, it was found that 
CV~Cyg (WDS J19543+3803) consisted of two almost identical stars separated by
$\rho = 0.7$~arcsec at $\theta = 140\degr$ and $\Delta V$ = 0.02.  
Surprisingly, though, our observations do not show any indications of 
a companion. Furthermore, the photometric study of \citet{vink1996} 
found no evidence for any third light in the system, and a single spectrum
of $H_\alpha$ taken by \citeauthor{vink1996} does not show any indication 
of triplicity.  We thus suspect the Hipparcos discovery is spurious.
\item[V2082~Cyg.]  A late-type companion to V2082~Cyg, with a light
contribution of $L_3/ (L_1 + L_2) \approx 0.02$ and a temperature
$T_3\simeq5100~$K, was found in Paper~II.  Independently, in Paper~I
we found an indication for a late-type tertiary from the relatively
high X-ray to bolometric flux ratio, which was unexpected for the
binary itself, since it has a moderately long orbital period, $P =
0.714$ days, and an early spectral type, F0 \citep{ddo9}.  At
$\rho=1.05$ arcsec, the system is very probably 
physically bound. The visual and infrared magnitudes are consistent
with a companion of the K2/3V spectral type.
\item[V829~Her.]  The possible multiplicity of V829~Her was indicated
by a large proper-motion error and an acceptable light-time effect
(LITE) solution for the contact binary (Paper~I). If one uses the
photometrically estimated parallax of \citet{bilir05}, $13.53 \pm
0.54$ mas with the LITE-derived $a_{12} \sin i = 0.9 \pm 0.2$ AU, and
assumes masses of components from \citet{priruc06}, the angular
separation should be about 0.08 arcsec. Thus, it is clear that the
observed LITE cannot be caused the object we detected at $\rho =
1.46$~arcsec.  If both the LITE and AO companions are confirmed, the
system would be part of a quadruple system.
\item[SW~Lac.] Indications for multiplicity of SW~Lac were found both
from spectra, which showed a late-type contribution
\citep{hendry1998}, and from complicated changes of its orbital period
\citep{prib1999}.  Our AO observations
show a relatively distant companion, at a separation of 1.68 arcsec.
Its physical bond to the contact binary is practically certain as
SW~Lac moved by 0.64~arcsec on the sky between 1998 and
2005 due to the fast proper motion, while the relative position
of the visual components has remained stable to within 0.01 arcsec.
The colors of the companion, $(H-K) = 0.11 - 0.16$, correspond to a
K5 -- M1V dwarf, which is consistent with the late K dwarf inferred
from the K-band magnitudes.   This visual companion might, despite the
large separation, be responsible for the spectral signature, but
cannot be responsible for the observed LITE.  Thus, another companion
could be hiding at still smaller separations. 
\item[XY~Leo.]  The multiple nature of XY~Leo was first indicated by
the light-time effect, and the interpretation and expected
nature of the third body was extensively discussed by
\citet{gehl72}. \citet{bard87} found spectroscopically that the
companion was a BY~Dra binary of a mid-M spectral type 
with a short orbital period of 
0.805 days. Thus, the system is a quadruple consisting of two
close binaries.  From the mass function of the third component
determined from the LITE orbit, \citet{bard87} estimated that the
orbital inclination of the outer orbit must be close to $90\degr$. The
visual pair has not yet been resolved directly.  If our marginal detection
(Section~\ref{close}) is confirmed, then -- 
when combined with the LITE parameters of \citet{priruc06}
-- the inclination of the outer
orbit should be about $67\degr$ with the longitude of the ascending node
of about $22\degr$ (or $202\degr$). The largest separation of
$\approx$ 0.17 arcsec appears to have occurred in the summer of 2003.
In 2013 occurs the second maximum at a separation of 0.13 arcsec.
Two or three spectroscopic runs within
the orbital period of the outer binary of 20 years would be needed to
lift the $\pm 180\degr$ ambiguity in the orientation of the orbit
($\Omega$) and to estimate the total mass of the whole quadruple system.
\item[V508~Oph.]
Multiplicity of V508~Oph was first indicated by the
Hipparcos astrometry, the system having an ``S'' flag
in H61 field. The estimated color of the visual
companion to V508~Oph, $(H-K)$ = 0.34, indicates a M5V
dwarf, consistent with the inferred absolute magnitude.
\item[V2388~Oph.]
V2388~Oph (WDS J17543+1108) is a known close
visual pair on a 8.92-year orbit with one component being
the eclipsing binary. The separation of the components
is only about 0.09 arcsec and the system has been a subject of
many speckle interferometry observations. The visual
magnitude difference of components was determined spectroscopically to
be  $L_3/(L_1 + L_2) = 0.20 \pm 0.02$, corresponding to
$\Delta V = 1.75\pm 0.02$~mag \citep{ddo6}.  Given the F2V classification
of V2388~Oph, the observed $\Delta V$ implies a G5 spectral type
for the companion.  Thus, one expects $\Delta K \approx = 0.8$, which
is consistent with what we find.  Oddly, however, the position
predicted by elements in Sixth Catalog of Orbits of Visual Binary
Stars \citep{hart2001}, $\rho=0.075$ arcsec and $\theta = 203\degr$,
is inconsistent with our observation.  Our measured
$\theta=31\pm13\degr$ suggests the components were swapped in the Sixth
Catalog. 
\item[ER~Ori.]  The visual companion to ER~Ori (WDS J05112-0833), at
$\rho = 0.187$~arcsec, $\theta = 354.4\degr$ and $\Delta V = 2.0$~mag
was first noticed in observations in March 1993 by \citet{goec94}.  In
our January 1998 observations, we find the companion at practically
the same position ($\rho = 0.183$~arcsec, $\theta = 354.4\degr$,
$\Delta K = 2.14$~mag), but, curiously, in our October 2005
observations there is no trace of it. Given the excellent detection
quality, the companion is almost certainly real, and almost 
certainly physically associated with the contact binary.
We note that 
the proper motion of the contact binary is directed to the South, while the
companion was observed to the North of the contact binary in 1993 and
1998; thus, if the stars are unrelated, the separation should have increased. 

The presence of a companion to the binary is also indicated by two
other effects.  First, there are large acceleration terms in the
Hipparcos astrometric solution, $g_\alpha =
-19.26\pm6.57{\rm\,mas\,yr^{-2}}$ and $g_\delta =
-17.34\pm4.71{\rm\,mas\,yr^{-2}}$, which suggest an orbital period of
only a few years.  Second, arrival times of the ER~Ori eclipsing
system clearly show the LITE, with an implied outer orbit with a period
of about 50 years \citep{kim03}, a semi-major axis for the eclipsing
pair of 6.7 AU, and substantial eccentricity, $e = 0.89$.  From this
orbit, periastron passage was predicted around July/August 2004, and
this might explain the absence of the visual companion during CFHT
2005 observing run. 
\item[U~Peg.] Although the visual companion to U~Peg is rather faint
and distant, the fast proper motion of the contact binary, amounting
to 0.515 arcsec between the 1998 and 2005 runs, has helped confirm the
physical association: the relative position of the components remained
stable within 0.03 arcsec.
\item[V592~Per.]  This binary (WDS J04445+3953) has a known close
visual companion that shows a slow orbital motion.  According to the WDS,
between 1977 and 2003, the PA changed from $\theta = 190$ to
$209\degr$. Our PA, $207 \pm 2\degr$ is consistent with the most
recent published observation in the WDS. The magnitude difference of
the components as given in \citet{wds} is $\Delta V$ = 0.85.  From our
AO observations, we derive $\Delta K = 0.39 \pm 0.06$, which is fairly
consistent with the estimated spectral type of the contact binary (F5-6, 
see \citet{ddo10}) and a G0V spectral type for the visual companion.
The components have very similar radial velocity \citep{ddo10}.
\item[CW~Sge.]  The binary was suspected to be a member of a
short-period visual pair because it is flagged by ``X'' (stochastic
astrometric solution) in the H59 Hipparcos catalog field (Paper~I).
This resulted in a Hipparcos parallax with a large error, $\pi = 2.57
\pm 4.14$ mas.  The Hipparcos so-called ``cosmic error'', $\epsilon =
7.64 \pm 1.32$ mas, suggests a rather small size of the astrometric
orbit.  The companion detected by our AO observations at a separation
of 1.84 arcsec cannot be identified with the one causing the rapid
astrometric motion.  Therefore CW~Sge may be a member of a system with
higher multiplicity.  

\item[BE~Scl.]  Multiplicity of BE~Scl was indicated by the large
``cosmic error'', $\epsilon = 13.36 \pm 1.35$ mas, in its Hipparcos
astrometric solution.  Our analysis shows a close companion at $\rho =
0.1$ arcsec with $\Delta K = 1.1\dots1.4$. This companion is likely
responsible for the observed astrometric motion.  With the large
photometric amplitude of about 0.45 mag, the eclipsing pair is an easy
object for timing of the eclipses and is expected to show variations
on time scales shorter than 1 year. 
\item[RZ~Tau] The possible multiplicity of RZ~Tau was first indicated
by the Hipparcos astrometry, RZ~Tau having an ``S'' flag in the H61
field.  The physical association of the visual pair is supported by a
stable relative position of its component between the 1998 and 2005
observations, despite the fact that proper motion of the contact
binary caused it to move 0.21 arcsec on the sky.
\end{description}

\section{SUMMARY}
\label{summary}

We present the results of an AO search for companions of contact
binary stars conducted on four nights in 1998 and 2005, 
in an attempt to confirm their high apparent incidence
indicated by various approaches in Paper~I.  The preliminary results
of the 1998 observations were included in Papers~I and II; the new
observations, taken in 2005, have contributed additional six
companions to the magnitude-limited sample ($V_{max}= 10$). 
The 2005 observations covered the Fall part of the CFHT sky and were
much more consistent in terms of the object selection than the 1998
observations. 
Fig.~\ref{fig8} gives the updated distribution of
projected separations for all objects of our program.  Our new
discoveries fall into the range of relatively small separations within
$0.5 < \log a(AU) < 2.0$.

\begin{figure} 
\epsscale{1.00}
\plotone{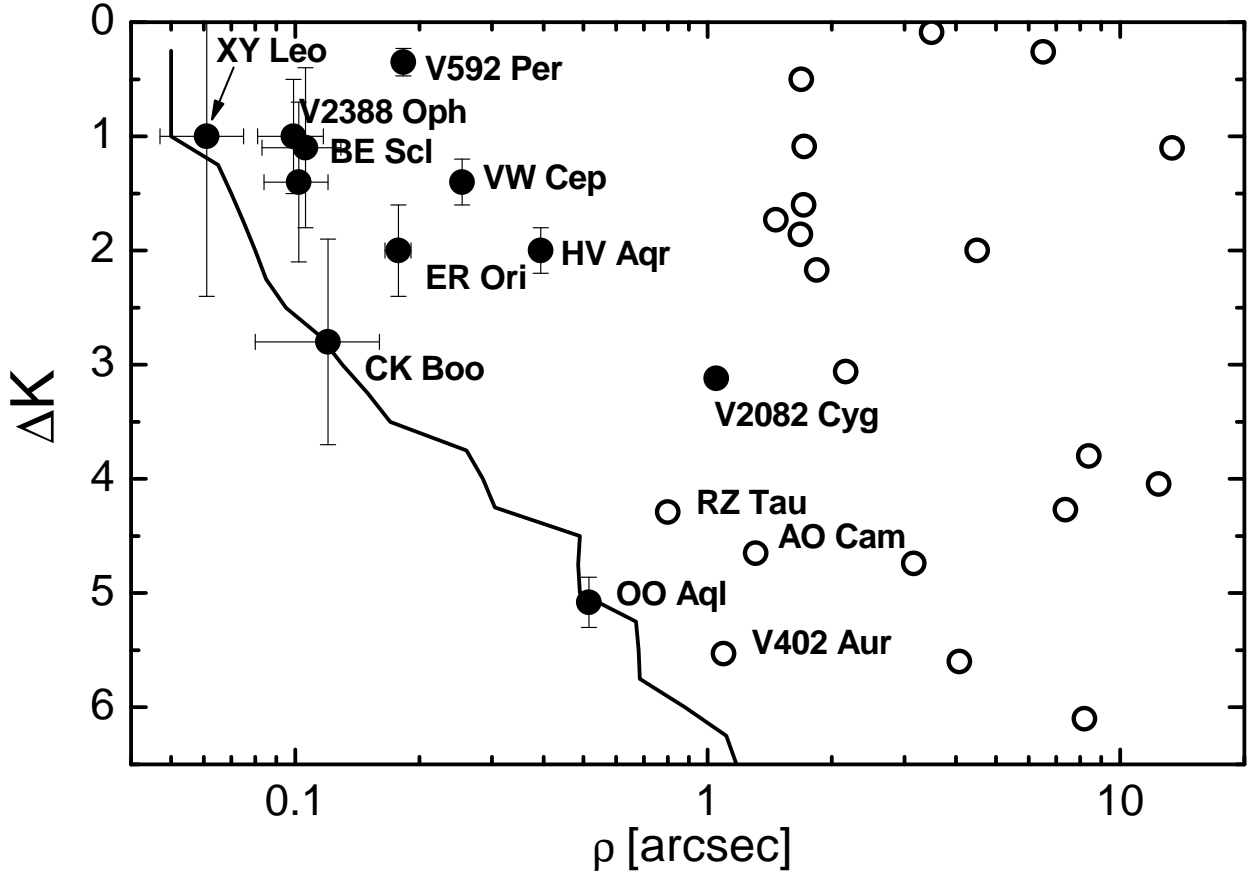}
\caption{Detections of companions in our program as a function of the
angular separation from the central star and the $K$ magnitude
difference. The detection limit evaluated by Monte Carlo experiments
is shown by the continuous curve. Detections resulting from our
automated search and modeling are shown by full circles while wider
pairs for which the measurement of the components was done separately
are shown by open circles.  Of the three sources near the detection
limit, the detection of the faint companion to OO~Aql is unambiguous,
but the cases of CK~Boo and XY~Leo require confirmation
(see Section~\ref{indiv} for details). \label{fig7}}
\end{figure}

\begin{figure} 
\epsscale{1.00}
\plotone{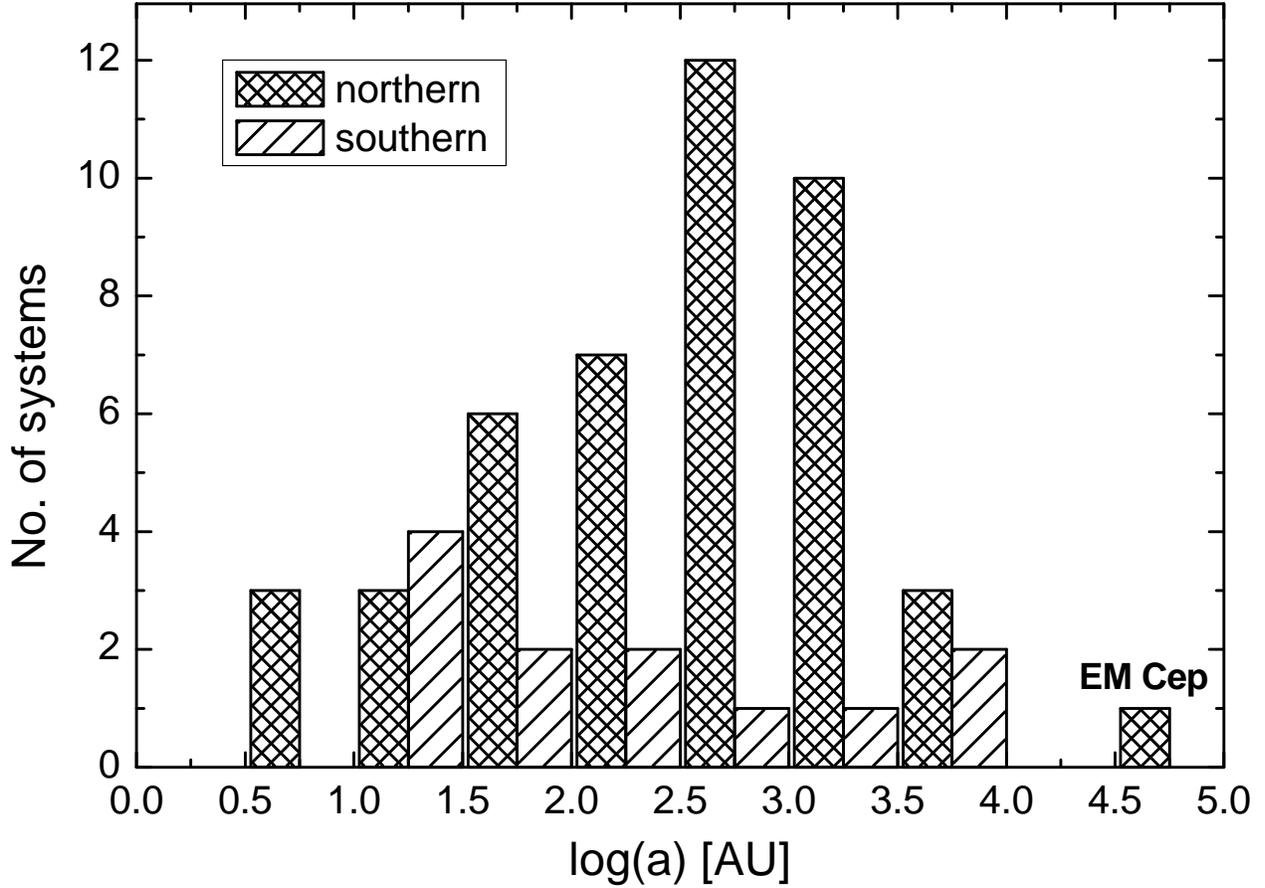}
\caption{Distributions of projected separations (in AU)
for resolved systems brighter than $V_{\rm max}=10$.  
This figure is an update to Fig.~9 in Paper~I
    \citep{priruc06}.  The distributions are shown separately for both
    hemispheres (relative to the equator), since the Northern
    hemisphere has been better studied.  The six new systems -- OO~Aql,
    HV~Aqr, V402~Aur, AO~Cam, V2082~Cyg, and XY~Leo -- are all located
    in bins between $0.5 < \log a(AU) < 2.0$.  \label{fig8}}
\end{figure}

For separations larger than one arcsec, the companions were easily
visible in the images, but for sub-arcsec detections we relied on an
automated search technique which was able to find companions hiding
within the AO diffraction rings.  This new technique permitted us to
approach the effective resolution limit of the CFHT AO system of about
0.07 -- 0.08 arcsec in the $K$-band.  The main results are shown in
Fig.~\ref{fig7} and are listed in Table~\ref{tab5}.  Thanks to this
new technique we were able to detect very close companions to XY~Leo,
V2388~Oph and BE~Scl, while for additional 
nine systems very close companions are
suspected and these systems require further observations. 
Especially encouraging are the AO detections of companions
previously indicated by other techniques in 
Papers~I and II (e.g., OO~Aql, V2082~Cyg, V829~Her, CW~Sge, BE~Scl).
We note that
contact binaries with third components with $\rho < 1''$ are perfect
objects for tests of ground-based interferometric system imaging
because a companion can provide an ideal phase
reference source.

From our detection limits shown in Fig.~\ref{fig7}, we can evaluate our
selection biases, and estimate how many systems might have been missed
by our AO observations at large magnitude differences $\Delta K$ and
small separations $\rho$.  From this figure, one sees that the
distribution of companions as a function of magnitude difference is
rather flat at large separations, $\rho > 1$ arcsec.  Similarly, for
relatively bright companions, one sees that the distribution in
$\log\rho$ is relatively flat.  If the two distributions are
independent, then this suggests that in addition to the eleven systems
at sub-arcsecond separations that we detected, another eleven or so
may have fallen below our separation detection limit 
(to $\Delta K \simeq 6$). 
If we corrected our results for those presumably missed 
companions, the implied incidence would increase from $(31 \pm 6)$\%
(25 detections in a sample of 80 contact binaries) to $(45 \pm 8)$\%.  

By combining the current results for the actual detections for binaries
with $V_{max}<10$ with our earlier AO detections reported
in Paper~I, we obtain the total fractional incidence of triple systems of 
($61 \pm 8$)\% for Northern objects
and ($19\pm5)$\% for Southern objects, strengthening the hypothesis
that all close binaries are members of multiple systems.
Finally, we note that from the 151 known contact binaries brighter than
$V_{max} = 10$~mag, so far only 51 have been observed with an AO system.
Extrapolating the detection rate from the present sample, one
would expect some 30 new AO detections or confirmations (of the cases
listed in Paper~I) for the one hundred binaries which remain to 
be observed with the AO technique. For the reason that still
about 2/3 of all systems remain to be observed, we defer a
detailed comparison of the current results with the statistics of
the solar-type field stars of \citet{DM91} until such a full 
sample becomes available.

\acknowledgements

The research made use of the SIMBAD database, operated at the CDS,
Strasbourg, France and accessible through the Canadian
Astronomy Data Centre, which is operated by the Herzberg Institute of
Astrophysics, National Research Council of Canada.
This research made also use of the Washington Double Star (WDS)
Catalog maintained at the U.S. Naval Observatory.
Support from the Natural Sciences and Engineering Council of Canada (NSERC)
to SMR and MHvK is acknowledged with gratitude; visits of TP to the 
University of Toronto were supported by the NSERC grant of SMR.

\clearpage

\clearpage

\clearpage

\clearpage

\clearpage

\clearpage

\clearpage

\clearpage

\clearpage

\clearpage

\begin{deluxetable}{cccl}

\tabletypesize{\footnotesize}

\tablewidth{0pt}
\tablenum{1}
\tablecolumns{4}

\tablecaption{Journal of the AO observations 
during the three observing runs of January 10, 1998,
July 23, 1998 and October 17--18, 2005. The full
table is available only in electronic form.
\label{tab1}}
\tablehead{
\colhead{Frame No.} &
\colhead{Object}  &
\colhead{HJD--2,400,000} &
Filter  \\
}
\startdata
419979  &  AQ~Psc  &  50824.7062  &  $K_{CO}$ \\
419980  &  AQ~Psc  &  50824.7081  &  $K_{CO}$ \\
419981  &  AQ~Psc  &  50824.7100  &  $K_{CO}$ \\
419982  &  AQ~Psc  &  50824.7122  &  $K_{CO}$ \\
419983  &  AQ~Psc  &  50824.7144  &  $K_{CO}$ \\
419984  &  AQ~Psc  &  50824.7165  &  $K_{CO}$ \\
419985  &  AQ~Psc  &  50824.7184  &  $K_{CO}$ \\
419986  &  AQ~Psc  &  50824.7203  &  $K_{CO}$ \\
419987  &  TT~Cet  &  50824.7291  &  $K$  \\
419988  &  TT~Cet  &  50824.7316  &  $K$  \\
\enddata
\tablecomments{ The first 10 rows of the table
are shown. The central wavelength and bandwidth of filters
used are as follows: \\
$K_{CO}$: $\lambda_{cen}$ = 2.298 $\mu$m, FWHM = 0.027 $\mu$m;\\
$K$: $\lambda_{cen}$ = 2.22  $\mu$m, FWHM = 0.4 $\mu$m;\\
$H$: $\lambda_{cen}$ = 1.65  $\mu$m, FWHM = 0.29  $\mu$m;\\
$H2$: $\lambda_{cen}$ = 2.122 $\mu$m, FWHM = 0.02  $\mu$m.\\
The $K_{CO}$ filter is entered as ``CO'' in the ASCII online version
of the table.
}
\end{deluxetable}

\begin{deluxetable}{lc|lc|lc}

\tabletypesize{\footnotesize}

\tablewidth{0pt}
\tablenum{2}
\tablecolumns{6}

\tablecaption{Overview of observed systems and detections of visual 
companions \label{tab2}}
\tablehead{
\colhead{Object} &
\colhead{Detection}  &
\colhead{Object} &
\colhead{Detection}  &
\colhead{Object} &
\colhead{Detection}  \\
}
\startdata
AB~And    & {\tt -NN} &  RW~Com    & {\tt N--} &  V753~Mon  & {\tt --N} \\
GZ~And    & {\tt -DD} &  RZ~Com    & {\tt N--} &  V502~Oph  & {\tt -N-} \\
V376~And  & {\tt --N} &  SX~Crv    & {\tt -N-} &  V508~Oph  & {\tt -D-} \\
EL~Aqr    & {\tt -N-} &  CV~Cyg    & {\tt --N} &  V566~Oph  & {\tt -NN} \\
HV~Aqr    & {\tt --D} &  DK~Cyg    & {\tt -NN} &  V839~Oph  & {\tt -N-} \\
OO~Aql    & {\tt --D} &  V401~Cyg  & {\tt -NN} &  V2388~Oph & {\tt --C} \\
V417~Aql  & {\tt -N-} &  V1073~Cyg & {\tt -NN} &  ER~Ori    & {\tt D-N} \\
V1464~Aql & {\tt --N} &  V2082~Cyg & {\tt --D} &  V1363~Ori & {\tt --N} \\
AH~Aur    & {\tt D-D} &  LS~Del    & {\tt --N} &  U~Peg     & {\tt -DD} \\
V402~Aur  & {\tt --D} &  SV~Equ    & {\tt -N-} &  BB~Peg    & {\tt -NN} \\
V410~Aur  & {\tt --D} &  UX~Eri    & {\tt N--} &  V335~Peg  & {\tt --N} \\
V449~Aur  & {\tt --N} &  YY~Eri    & {\tt N-N} &  V351~Peg  & {\tt --N} \\
TZ~Boo    & {\tt -N-} &  BV~Eri    & {\tt -NN} &  V357~Peg  & {\tt --N} \\
VW~Boo    & {\tt -N-} &  FP~Eri    & {\tt --N} &  KN~Per    & {\tt --N} \\
XY~Boo    & {\tt -N-} &  AK~Her    & {\tt -D-} &  V592~Per  & {\tt --D} \\
AC~Boo    & {\tt -N-} &  V829~Her  & {\tt --D} &  VZ~Psc    & {\tt -N-} \\
CK~Boo    & {\tt -C-} &  V972~Her  & {\tt --N} &  AQ~Psc    & {\tt NNN} \\
AO~Cam    & {\tt --D} &  FG~Hya    & {\tt N--} &  TY~Pup    & {\tt N--} \\
DN~Cam    & {\tt --N} &  SW~Lac    & {\tt -DD} &  CW~Sge    & {\tt --D} \\
BH~CMi    & {\tt --N} &  V407~Lac  & {\tt --N} &  BE~Scl    & {\tt --C} \\
V523~Cas  & {\tt --N} &  UZ~Leo    & {\tt N--} &  RZ~Tau    & {\tt DDD} \\
VW~Cep    & {\tt -D-} &  XY~Leo    & {\tt C--} &  EQ~Tau    & {\tt --N} \\
TT~Cet    & {\tt NN-} &  XZ~Leo    & {\tt N--} &  V781~Tau  & {\tt N-N} \\
CL~Cet    & {\tt --N} &  AP~Leo    & {\tt N--} &  AG~Vir    & {\tt N--} \\
CT~Cet    & {\tt --D} &  VZ~Lib    & {\tt -N-} &  AH~Vir    & {\tt D--} \\
DY~Cet    & {\tt --N} &  UV~Lyn    & {\tt N--} &  GR~Vir    & {\tt -N-} \\
RS~Col    & {\tt --N} &  V752~Mon  & {\tt --D} &            &           \\
\enddata
\tablecomments{Results of the three observing runs of January 10, 1998,
July 23, 1998 and October 17--18, 2005 are in coded in three
columns as follows:
              {\tt -} = not observed in a given run,
              {\tt N} = no detection,
              {\tt D} = detection at a separation of $<5$ arcsec,
              {\tt C} = a very close pair; possible
               detection through a deformation of the diffraction pattern.}
\end{deluxetable}

\begin{deluxetable}{lcrccrrcrrrc}

\tabletypesize{\scriptsize}
\tabcolsep 3pt
\tablecolumns{12}
\tablewidth{0pt}
\tablenum{3}
\tablecaption{Results of an automated search for close companions to
observed contact binaries: New detections and suspect cases \label{tab3}.}
\tablehead{
 \colhead{Name}            & 
 \colhead{Year}            & 
 \colhead{$N_f$}           & 
 \colhead{$N_t$}           & 
 \colhead{$N_b$}           & 
 \colhead{$C$}             & 
 \colhead{$C_n$}           & 
 \colhead{$r$}             & 
 \colhead{$S$}             & 
 \colhead{$B$}             & 
 \colhead{$R_n$}           & 
 \colhead{Flag}           \\ 
}
\startdata
V376 And  & 2005 &  10 &   46&  268&  11&   5.5&  4&   46&    84&   14.7& S  \\
HV Aqr    & 2005 &   5 &   46&  268&  51&  51.0&  4&  141&     4& 1839.4& D  \\
TZ Boo    & 1998 &   4 &   28&  176&   8&  16.4&  3&   25&    54&   20.8& S  \\
CK Boo    & 1998 &   8 &   28&  176&  13&  13.3&  5&  103&    15&  155.7& D  \\
DN Cam    & 2005 &   5 &   46&  268&  20&  20.0&  4&  154&   350&   23.6& S  \\
VW Cep    & 1998 &   4 &   28&  176&  63& 129.4&  3&  107&     4& 1209.4& D  \\
YY Eri    & 2005 &   5 &   46&  268&   9&   9.0&  4&   80&   830&    5.2& S  \\
BV Eri    & 2005 &   5 &   46&  268&  12&  12.0&  5&   89&   820&    5.8& S  \\
V829 Her  & 2005 &   5 &   46&  268&  11&  11.0&  6&  144&   597&   12.9& S  \\
XY Leo    & 1998 &   8 &   28&  176&  23&  23.6&  4&   98&   104&   20.9& D  \\
V508 Oph  & 1998 &   4 &   28&  176&   7&  14.4&  5&   58&   336&    7.6& S  \\
V2388 Oph & 2005 &   5 &   46&  268&  21&  21.0&  4&  144&   109&   70.8& D  \\
ER Ori    & 1998 &   8 &   28&  176&  41&  42.1&  4&  188&    44&   93.6& D  \\
V351 Peg  & 2005 &  10 &   46&  268&  13&   6.5&  4&   38&    54&   18.8& S  \\
V592 Per  & 2005 &   5 &   46&  268& 134& 134.0&  4&  221&    62&  192.2& D  \\
TY Pup    & 1998 &   8 &   28&  176&   8&   8.2&  6&  121&   358&    7.4& S  \\
BE Scl    & 2005 &  10 &   46&  268&  93&  46.5&  4&  372&   127&   78.4& D  \\
\enddata
\tablecomments{
Explanation of columns: Name -- Variable star name in the General Catalog of
Variable Stars; Year -- year of the observing run; $N_f$ -- number of object frames
used for the count map; $N_t$ -- number of templates used; $N_b$ -- number of
frames coming into the background map; $C$ -- maximum count for the object;
$C_n$ -- normalized maximum count (see text for the definition);
$r$ -- radius of the aperture for the determination of the integrated counts in the
object count map; $S$ -- sum of counts within the selected aperture in the object
map; $B$ -- corresponding sum of counts in the background map;
$R_n$ -- normalized ratio of the summed
counts in the object and the background map (see the text for the definition);
Flag -- flag indicating status of the detection: ``D'' -- detection, ``S'' --
suspected case.
The detection level is set at $C_n > 20$ or $R_n > 50$.
Only detections with automated search are taken into account; binaries
with directly visible components are listed in Table~2 (see also Fig.~\ref{fig3}).
}
\end{deluxetable}

\begin{deluxetable}{lcr|lcr|lcr}

\tabletypesize{\scriptsize}
\tabcolsep 3pt
\tablecolumns{9}
\tablewidth{0pt}
\tablenum{4}
\tablecaption{Results of an automatic search for close companions:
Contact binary systems not detected to have visual companions \label{tab4}}
\tablehead{
 \colhead{Name}            & 
 \colhead{Year}            & 
 \colhead{$C_n$}           & 
 \colhead{Name}            & 
 \colhead{Year}            & 
 \colhead{$C_n$}           & 
 \colhead{Name}            & 
 \colhead{Year}            & 
 \colhead{$C_n$}           \\ 
}
\startdata
AB And    &  1998 &  6.2  & DK Cyg    &  2005 &  2.0  & V566 Oph  &  2005 &  7.0  \\
AB And    &  2005 &  2.5  & V401 Cyg  &  1998 &  8.2  & V839 Oph  &  1998 & 14.4  \\
GZ And    &  1998 &  2.1  & V401 Cyg  &  2005 &  7.0  & ER Ori    &  2005 &  4.5  \\
GZ And    &  2005 &  9.0  & V1073 Cyg &  1998 & 12.3  & V1363 Ori &  2005 &  3.0  \\
EL Aqr    &  1998 &  8.2  & V1073 Cyg &  2005 &  7.0  & U Peg     &  1998 &  4.1  \\
AH Aur    &  2005 &  4.0  & V2082 Cyg &  2005 &  3.0  & U Peg     &  2005 &  2.0  \\
V449 Aur  &  2005 & 10.0  & LS Del    &  2005 &  4.0  & BB Peg    &  2005 &  6.0  \\
OO Aql    &  2005 &  4.0  & SV Equ    &  1998 & 16.4  & V335 Peg  &  2005 &  4.5  \\
V417 Aql  &  1998 &  8.2  & YY Eri    &  1998 &  6.2  & V357 Peg  &  2005 &  4.0  \\
V1464 Aql &  2005 &  6.0  & BV Eri    &  1998 &  2.1  & KN Per    &  2005 &  6.0  \\
V402 Aur  &  2005 &  1.5  & FP Eri    &  2005 &  2.5  & VZ Psc    &  1998 & 10.3  \\
V410 Aur  &  2005 &  3.0  & AK Her    &  1998 &  4.1  & AQ Psc    &  1998 &  3.1  \\
VW Boo    &  1998 &  2.1  & V972 Her  &  2005 & 10.0  & AQ Psc    &  1998 &  2.1  \\
AC Boo    &  1998 &  4.1  & SW Lac    &  1998 & 10.3  & AQ Psc    &  2005 &  5.0  \\
AO Cam    &  2005 &  7.0  & SW Lac    &  2005 &  7.0  & CW Sge    &  2005 &  4.0  \\
BH CMi    &  2005 &  6.0  & V407 Lac  &  2005 &  9.0  & EQ Tau    &  2005 &  2.0  \\
V523 Cas  &  2005 &  2.0  & XZ Leo    &  1998 &  4.1  & RZ Tau    &  1998 &  2.1  \\
CL Cet    &  2005 &  4.5  & AP Leo    &  1998 &  2.1  & RZ Tau    &  2005 &  3.0  \\
CT Cet    &  2005 &  9.0  & VZ Lib    &  1998 &  4.1  & V781 Tau  &  1998 &  3.1  \\
DY Cet    &  2005 &  4.5  & UV Lyn    &  1998 &  6.2  & V781 Tau  &  2005 &  3.0  \\
RS Col    &  2005 & 11.0  & V752 Mon  &  2005 &  5.0  & AG Vir    &  1998 &  2.1  \\
RZ Com    &  1998 &  3.1  & V753 Mon  &  2005 &  5.0  & AH Vir    &  1998 &  6.2  \\
SX Crv    &  1998 &  4.1  & V502 Oph  &  1998 &  6.2  & GR Vir    &  1998 &  3.1  \\
CV Cyg    &  2005 &  2.0  & V566 Oph  &  1998 & 10.3  &           &       &       \\
\enddata
\tablecomments{
Explanation of columns: Name -- Variable star name in the General Catalog of
Variable Stars; Year -- year of the observing run; $C_n$ -- normalized maximum
count (see text for the definition).\\
Note: AQ~Psc was observed both in January and July 1998.
}
\end{deluxetable}


\begin{deluxetable}{lcrrrrrccccl}

\tabletypesize{\footnotesize}

\tabcolsep 3pt

\tablecolumns{12}

\tablewidth{0pt}
\tablenum{5}

\tablecaption{Companions of contact binaries detected or confirmed
during this program \label{tab5}}
\tablehead{
\colhead{Name}            & 
\colhead{HJD}             & 
\colhead{Phase}           & 
\colhead{$\Delta V_\phi$}     & 
\colhead{$\mu \Delta t$}  & 
\colhead{$\rho$}          & 
\colhead{$\theta$}        & 
\colhead{$\Delta K$}      & 
\colhead{$\Delta H$}      & 
\colhead{$M_K$}           & 
\colhead{$M_H$}           & 
\colhead{Sp.}             \\ 
                          & 
\colhead{2\,400\,000+}    & 
                          & 
                          & 
\colhead{[mas]}           & 
\colhead{[arcsec]}        & 
\colhead{[deg]}           & 
                          & 
                          & 
                          & 
                          & 
                          \\} 
\startdata
\sidehead{\bf New detections}
GZ And   &  51019.1015&   0.484&  0.66&    0.0& 2.130(6)  &   33.55(7)  &  2.45(5)   &                &  5.99 &       & M3V \\
GZ And   &  51019.1102&   0.513&  0.67&    0.0& 2.131(11) &   33.51(10) &            &   2.625(9)     &       &  6.24 & M3-4\\
GZ And   &  53660.9608&   0.815&  0.04&  116.5& 2.163(7)  &   29.25(13) &  3.069(11) &                &  5.99 &       & M3V \\
HV Aqr   &  53660.7804&   0.315&  0.04&    0.0& 0.394(9)  &   12.4(8)   &  2.0(3)    &                &  4.3  &       & K2-3V\\
OO Aql   &  53660.7339&   0.835&  0.16&    0.0& 0.510(16) &  290(2)     &  5.1(3)    &                &  7.2  &       & $>$M5V\\
AH Aur   &  50824.9048&   0.072&  0.29&    0.0& 3.189(4)  &   62.91(13) &  4.31(4)   &                &  6.30 &       & M5V  \\
AH Aur   &  50824.9169&   0.097&  0.21&    0.0& 3.188(6)  &   62.98(18) &            &   4.31(4)      &       &  6.26 & M3-4V\\
AH Aur   &  53662.1136&   0.170&  0.06&  154.0& 3.159(4)  &   57.3(6)   &  4.74(5)   &                &  6.49 &       & $>$M5V\\
V402 Aur &  53661.0890&   0.990&  0.13&    0.0& 1.093(3)  &   47.1(3)   &  5.53(24)  &                &  7.00 &       & $>$M5V\\
V402 Aur &  53662.0200&   0.533&  0.12&    0.0& 1.093(3)  &   47.0(2)   &  5.29(3)   &                &  6.74 &       & $>$M5V\\
CK Boo   &  51018.8040&   0.351&  0.08&    0.0& 0.12(4)   &  198(5)     &  2.8(9)    &                &  5.3  &       & M1V \\
AO Cam   &  53662.0455&   0.057&  0.43&    0.0& 1.309(5)  &   87.87(15) &  4.65(7)   &                &  7.68 &       & $>$M5V\\
V2082 Cyg&  53660.7223&   0.927&  0.06&    0.0& 1.049(4)  &  336.92(19) &  3.12(4)   &                &  4.32 &       & K2V \\
V829 Her &  53661.7171&   0.190&  0.03&    0.0& 1.463(9)  &  344.69(22) &  1.726(24) &                &  4.07 &       & K1V \\
SW Lac   &  51019.0182&   0.825&  0.10&    0.0& 1.680(1)  &   85.33(9)  &  2.554(9)  &                &  5.03 &       & K9V \\
SW Lac   &  51019.0251&   0.847&  0.14&    0.0& 1.679(2)  &   85.36(19) &            &   2.612(7)     &       &  5.19 & K9V \\
SW Lac   &  53660.8173&   0.004&  0.84&  642.3& 1.680(4)  &   79.16(4)  &  1.86(3)   &                &  5.08 &       & M0V \\
XY Leo   &  50825.0225&   0.915&  0.20&    0.0& 0.061(14) &   91(7)     &  1.0(13)   &                &  4.7  &       & K6V \\
V508 Oph &  51018.9071&   0.206&  0.02&    0.0& 2.398(3)  &   18.85(5)  &  3.968(11) &                &  6.50 &       & $>$M5V\\
V508 Oph &  51018.9152&   0.229&  0.00&    0.0& 2.395(3)  &   18.89(9)  &            &   4.275(11)    &       &  6.84 & $>$M5V\\
U Peg    &  51019.0677&   0.149&  0.10&    0.0& 4.052(5)  &  275.38(9)  &  5.232(21) &                &  7.78 &       & $>$M5V\\
U Peg    &  53660.8890&   0.206&  0.02&  515.0& 4.076(5)  &  271.21(11) &  5.60(5)   &                &  8.07 &       & $>$M5V\\
U Peg    &  53661.8843&   0.861&  0.12&  515.2& 4.080(4)  &  271.25(11) &  5.51(4)   &                &  8.09 &       & $>$M5V\\
CW Sge   &  53660.7637&   0.907&  0.19&    0.0& 1.838(3)  &   60.68(17) &  2.17(3)   &                &  3.51 &       & G5V \\
BE Scl   &  53660.9444&   0.059&  0.33&    0.0& 0.102(19) &   72(4)     &  1.4(7)    &                &  3.8  &       & G8V \\
BE Scl   &  53661.9577&   0.455&  0.35&    0.0& 0.106(23) &   74(4)     &  1.1(7)    &                &  3.4  &       & G5V \\
RZ Tau   &  50824.8140&   0.033&  0.60&    0.0& 0.796(10) &   43.76(10) &  3.61(4)   &                &  6.23 &       & M5V \\
RZ Tau   &  50824.8249&   0.059&  0.45&    0.0& 0.794(6)  &   43.7(4)   &            &   3.60(6)      &       &  6.10 & M3V \\
RZ Tau   &  51019.1256&   0.490&  0.58&   14.4& 0.799(2)  &   44.02(11) &  3.71(3)   &                &  6.31 &       & $>$M5V\\
RZ Tau   &  53662.0090&   0.503&  0.58&  210.5& 0.801(4)  &   38.93(11) &  4.29(5)   &                &  6.89 &       & $>$M5V\\
\sidehead{\bf Confirmed detections}
V410 Aur &  53661.0756&   0.244&  0.03&    0.0& 1.716(6)  &  224.59(10) &  1.091(15) &                &  3.52 &       & K2V \\
VW Cep   &  51018.9527&   0.326&  0.06&    0.0& 0.254(8)  &  166.8(20)  &  1.4(2)    &                &       &       & K4V \\
CT Cet   &  53661.9190&   0.795&  0.05&    0.0& 3.492(8)  &  206.73(8)  &  0.09(4)   &                &       &       & G3V \\
AK Her   &  51018.8769&   0.615&  0.14&    0.0& 4.459(5)  &  324.37(7)  &  2.019(9)  &                &  4.13 &       & K2V \\
V752 Mon &  53661.1142&   0.862&  0.01&    0.0& 1.686(4)  &   23.89(15) & -0.50(2)   &                &       &       &     \\
V2388 Oph&  53661.7303&   0.531&  0.22&    0.0& 0.099(18) &   31(13)    &  1.0(5)    &                &       &       & F0V \\
ER Ori   &  50824.8540&   0.197&  0.03&    0.0& 0.178(14) &  356(5)     &  2.0(4)    &                &       &       & K2V \\
ER Ori   &  53661.1312&   0.981&  0.63&  207.3&           &             &            &                &       &       & K2V \\
V592 Per &  53661.0618&   0.675&      &    0.0& 0.183(3)  &  207.9(14)  &  0.39(12)  &                &       &       & G0: \\
AH Vir   &  50825.1084&   0.380&  0.15&    0.0& 1.707(2)  &   16.50(7)  &  1.551(21) &                &  3.98 &       & K1V \\
\enddata
\tablecomments{Explanation of columns: Name = Variable star name
in the General Catalog of Variable Stars;
HJD = Heliocentric Julian date of the particular AO observation;
Phase = Orbital phase of the binary calculated from the ephemeris given below;
$\Delta V_\phi$ = $V_{\rm obs} - V_{\rm max}$: Correction
required to bring the magnitude to the maximum visual brightness of
the eclipsing pair for the instant of observation (see text);
$\mu \Delta t$ = Cumulative proper motion of the binary
counted from the first observation (always zero for
stars observed during a single night);
$\rho$ = Angular separation of the components;
$\theta$ = Position angle of the secondary (fainter) component;
$\Delta K$ and $\Delta H$: Measured magnitude differences between
the visual companion and the binary (without correction for the phase of the
eclipsing pair);
$M_K$ and $M_H$: Absolute magnitude of the companion
determined from the estimated absolute
$K$ and $H$ magnitude of the contact binary (Table~\ref{tab4}),
$\Delta K$ and $\Delta H$ and $\Delta V_\phi$
(see text); Sp.: estimated spectral type of the visual companion.
\\[1ex]
Ephemerides ($HJD_{min} - $2,400,000 + period in days) used for the computation of
phases: \\
 GZ~And:    52500.1198 + 0.3050177;   HV~Aqr:    52500.2163 + 0.3744582; \\
 OO~Aql:    52500.261  + 0.5067932;   AH~Aur:    52500.3848 + 0.4941067; \\
 V402~Aur:  52500.567  + 0.60349867;  V410~Aur:  52500.0033 + 0.3663562; \\
 CK~Boo:    52500.026  + 0.3551538;   AO~Cam:    52500.1061 + 0.3299036; \\
 VW~Cep:    52500.0321 + 0.2783108;   CT~Cet:    48500.1847 + 0.2564863; \\
 V2082~Cyg: 52466.1122 + 0.714084;    AK~Her:    52500.2709 + 0.421523;  \\
 V829~Her:  52500.159  + 0.358153;    SW~Lac:    52500.1431 + 0.3207165; \\
 XY~Leo:    52500.0872 + 0.2840978;   V752~Mon:  48500.2837 + 0.462902;  \\
 V508~Oph:  52500.0545 + 0.3447901;   V2388~Oph: 52500.379  + 0.8022979; \\
 ER~Ori:    52500.1715 + 0.4234018;   U~Peg:     52500.1288 + 0.3747766;  \\
 V592~Per   53399.3400 + 0.715722;    CW~Sge:    52500.567  + 0.6603631; \\
 BE~Scl:    52500.0549 + 0.42290144;  RZ~Tau:    52500.3968 + 0.4156776; \\
 AH~Vir:    52500.3174 + 0.407532;                                       \\
}
\end{deluxetable}

\begin{deluxetable}{lrrrrrcccccccccccrr}

\tabletypesize{\scriptsize}

\rotate

\tabcolsep 3pt

\tablecolumns{19}

\tablewidth{0pt}
\tablenum{6}

\tablecaption{Properties of contact binaries with
newly detected companions \label{tab6}}

\tablehead{
\colhead{Name}            & 
\colhead{$V_{\rm max}$}   & 
\colhead{$\Delta V$}      & 
\colhead{$\pi$}           & 
\colhead{$\sigma_\pi$}    & 
\colhead{$\mu$}           & 
\colhead{Sp. type}        & 
\colhead{$(B-V)_0$}       & 
\colhead{$M_V$}           & 
\colhead{$M_V$}           & 
\colhead{$M_H$}           & 
\colhead{$M_K$}           & 
\colhead{$M_H$}           & 
\colhead{$M_K$}           & 
\colhead{$M_{12}$}        & 
\colhead{Sp. type}        & 
\colhead{$M_{3}$}         & 
\colhead{$a$}             & 
\colhead{$P_{vis}$}       \\ 
                          & 
                          & 
                          & 
\colhead{[mas]}           & 
\colhead{[mas]}           & 
\colhead{[${\rm mas\,yr^{-1}}$]}         & 
\colhead{(12)}            & 
                          & 
\colhead{calc}            & 
\colhead{spec}            & 
\colhead{spec}            & 
\colhead{spec}            & 
\colhead{corr}            & 
\colhead{corr}            & 
\colhead{[$M_\odot$]}     & 
\colhead{comp.}           & 
\colhead{[$M_\odot$]}     & 
\colhead{[AU]}            & 
\colhead{[yr]}            \\}
\startdata
GZ~And    & 10.83 & 0.78 &  5.33 &      &  16.1 & G5V   & 0.68 & 4.46 & 5.10 & 3.52 & 3.58 &  2.94 & 2.88 & 1.708 & M3-4V   &  0.3    & 400 &   5,500\\
HV~Aqr    &  9.71 & 0.40 &  5.33 &      & 112.0 & F5V   & 0.44 & 3.34 & 3.50 & 2.40 & 2.44 &  2.28 & 2.24 & 1.569 & K2-3V   &  0.7    &  74 &     430\\
OO~Aql    &  9.20 & 0.80 &  7.19 &      &  66.2 & G5V   & 0.68 & 3.48 & 5.10 & 3.52 & 3.58 &  1.96 & 1.90 & 1.954 & $>$M5V  &  $<$0.2 &  71 &     420\\
AH~Aur    & 10.20 & 0.37 &  6.18 & 2.05 &  19.8 & F7V   & 0.49 & 2.96 & 3.83 & 2.57 & 2.61 &  1.74 & 1.70 & 1.967 & M5V     &  0.2    & 510 &   8,200\\
V402~Aur  &  8.84 & 0.14 &  7.01 & 1.31 &  11.3 & F2V   & 0.35 & 2.15 & 3.60 & 2.78 & 2.82 &  1.37 & 1.33 & 1.965 & $>$M5V  &  $<$0.2 & 156 &   1,370\\
CK~Boo    &  8.99 & 0.27 &  6.38 & 1.34 & 111.9 & F7.5V & 0.51 & 3.66 & 3.92 & 2.61 & 2.64 &  2.37 & 2.35 & 1.569 & M1V     &  0.45   &  19 &      54\\
AO~Cam    &  9.50 & 0.50 &  7.98 &      &   0.0 & G0V   & 0.58 & 4.01 & 4.40 & 2.99 & 3.04 &  2.65 & 2.60 & 1.605 & $>$M5V  &  $<$0.2 & 164 &   1,630\\
V2082~Cyg &  6.63 & 0.05 & 11.04 & 0.56 &  97.0 & F0V   & 0.30 & 1.84 & 2.70 & 2.00 & 2.03 &  1.17 & 1.14 &       & K2V     &  0.75   &  95 &        \\
V829~Her  & 10.10 & 0.29 &  4.97 &      &  19.6 & F7V   & 0.49 & 3.58 & 3.83 & 2.57 & 2.61 &  2.36 & 2.32 & 1.806 & K1V     &  0.77   & 295 &   3,570\\
SW~Lac    &  8.51 & 0.88 & 12.30 & 1.26 &  88.8 & G5V   & 0.68 & 3.96 & 5.10 & 3.52 & 3.58 &  2.44 & 2.38 & 2.204 & K9V     &  0.54   & 136 &     940\\
XY~Leo    &  9.45 & 0.48 & 15.86 & 1.80 &  81.3 & K0V   & 0.81 & 5.45 & 5.90 & 3.94 & 4.02 &  3.57 & 3.49 & 1.188 & K6V     &  0.64   & 3.8 &     5.5\\
V508~Oph  & 10.06 & 0.63 &  7.68 & 2.14 &  47.4 & G0V   & 0.58 & 3.92 & 4.40 & 2.99 & 3.04 &  2.56 & 2.51 & 1.520 & $>$M5V  &  $<$0.2 & 310 &   4,400\\
U~Peg     &  9.23 & 0.84 &  7.18 & 1.43 &  71.2 & G2V   & 0.63 & 3.92 & 4.70 & 3.24 & 3.29 &  2.51 & 2.46 & 1.554 & $>$M5V  & $<$0.2  & 560 &  10,600\\
CW~Sge    & 11.13 & 0.80 &  2.57 & 4.14 &   2.1 & F5V   & 0.44 & 2.25 & 3.50 & 2.40 & 2.44 &  1.19 & 1.15 &       & G5V     &   0.92  & 715 &        \\
BE~Scl    & 10.24 & 0.43 &  9.76 & 5.11 &  19.5 & F8V   & 0.52 & 3.35 & 4.00 & 2.65 & 2.70 &  2.05 & 2.00 &       & G5-8V   & 0.8-0.9 & 10  &        \\
RZ~Tau    & 10.08 & 0.63 &  5.74 & 1.85 &  27.1 & F0V   & 0.30 & 2.72 & 2.70 & 2.00 & 2.03 &  2.05 & 2.02 &       & $>$M5V  &  $<$0.2 & 139 &        \\
\enddata
\tablecomments{
Explanation of columns: Name = Variable star name
General Catalog of Variable Stars;
$V_{\rm max}$: Johnson or transformed Hipparcos $H_p$ maximum light magnitude;
$\Delta V$: Amplitude of the light variations;
$\pi$ and $\sigma_\pi$: Trigonometric parallax  and its error (if
given without error, photometrically determined parallax; see text);
$\mu = \sqrt{(\mu_\alpha \cos \delta)^2 + (\mu_\delta)^2}$: Proper motion;
Sp. type (12): Spectral type of the binary;
$(B-V)_0$: the de-reddened color index;
$M_V$ (calc): Absolute visual magnitude calculated from
$V_{max}$ and Hipparcos parallax (for V2082~Cyg, SW~Lac, and XY~Leo)
or from the \citet{rd97} calibration;
$M_{V,H,K}$ (spec): Absolute $V,H,K$ magnitudes corresponding
to a main sequence star of the same spectral type as the contact binary;
$M_{H,K}$ (corr): Corrected absolute magnitudes of the contact pair (see text);
$M_{12}$: Total (not projected, $\sin i$ accounted for) mass of the contact binary;
Sp. type comp.: estimated spectral type of the visual companion;
$M_{3}$: estimated mass of the visual companion (according to tabulation in \citet{cox00})];
$a$: Projected separation in astronomical units determined from angular separation
and distance; $P_{vis}$: Estimated period of the visual orbit.
}
\end{deluxetable}
\end{document}